\documentclass[12pt]{iopart}

\expandafter\let\csname equation*\endcsname\relax
\expandafter\let\csname endequation*\endcsname\relax
\usepackage{amsmath}
\usepackage{iopams}
\usepackage{dcolumn}
\usepackage{bm}
\usepackage{graphicx}
\usepackage{multirow,setspace,times,graphicx,color,rotating,subfigure,url}

\begin{document}

\title[Information Diffusion Backbones in Temporal Networks]{Information diffusion backbones in temporal networks}

\author{Xiu-Xiu Zhan, Alan Hanjalic, Huijuan Wang$^{1}$}

\address{$^1$ Faculty of Electrical Engineering, Mathematics, and Computer Science, Delft University of Technology, Delft, 2628CD, The Netherlands }

\ead{H.Wang@tudelft.nl}

\begin{abstract}
Information diffusion on a temporal network can be modeled by the Susceptible-Infected ($SI$) spreading process. An infected (information possessing) node could spread the information to a Susceptible node with a given spreading probability $\beta$ whenever a contact happens between the two nodes. Progress has been made in understanding how temporal network features affect the percentage of nodes reached by the information. In this work, we explore further: which node pairs are likely to contribute to the actual diffusion of information, i.e. appear in a diffusion trajectory? How is this likelyhood related to the local temporal connection features of the node pair? Such deep understanding of the role of node pairs is crucial to explain and control the prevalence of information spread. We consider a large number of real-world temporal networks. First, we propose the construction of an \emph{information diffusion backbone} $G_{B}(\beta)$ for a $SI$ spreading process with an infection probability $\beta$ on a temporal network. The backbone is a weighted network where the weight of each node pair indicates how likely the node pair appear in a diffusion trajectory starting from an arbitrary node. Second, we investigate the relation between the backbones with different infection probabilities on a temporal network. We find that the backbone topologies obtained for low and high infection probabilities approach the backbone $G_{B}(\beta\to 0)$ and $G_{B}(\beta=1)$, respectively. The backbone $G_{B}(\beta\to0)$ equals the integrated weighted network, where the weight of a node pair counts the total number of contacts in between. Finally, we discover a local connection feature among many other features that could well predict the links in $G_{B}(\beta=1)$, whose computation complexity is high. This local feature encodes the time that each contact occurs, pointing out the importance of temporal features in determining the role of node pairs in a dynamic process.

\end{abstract}

\maketitle

\section{Introduction}
Both online social networks like Facebook, Twitter and LinkedIn and physical contact networks facilitate the diffusion of information where a piece of information is transmitted from one individual to another through their online or physical contacts or interactions. Information diffusion processes have been modeled by e.g. independent cascade models~\cite{watts2002simple}, threshold models~\cite{granovetter1978threshold} and epidemic spreading models~\cite{pastor2015epidemic, liu2015events,zhang2016dynamics, epi_inter,RevModPhys.87.925,heter_bo}. Social networks have been first considered to be static where nodes represent the individuals and where links indicate the relationship between nodes such as whether they have ever contacted or not~\cite{barabasi2016network}. Information is assumed to propagate through the static links according to the aforementioned models. Recently, the temporal nature of contact networks have been taken into account in the spreading processes, i.e. the contacts between a node pair occur at specific time stamps (the link between nodes is time dependent)  and information could possibly propagate only through contacts (or temporal links) ~\cite{holme2012temporal, holme2015modern, scholtes2014causality, valdano2015analytical,zhang2017spectral}. Consider the $SI$ (Susceptible-Infected) spreading process on a temporal network~\cite{pastor2015epidemic,zhang2016dynamics}. Each individual can be in one of the two states: susceptible ($S$) or infected ($I$). A node in the infected (susceptible) state means that it has (does not have) the information. A susceptible node could get infected with an infection probability $\beta$ via each contact with an infected node. An infected individual remains infected forever.

Progress has been made in the exploration of how temporal network features~\cite{karsai2011small,lambiotte2013burstiness, moinet2015burstiness,hethcote2000mathematics, rocha2013bursts} and the choice of the source node~\cite{lee2012exploiting, starnini2013immunization} influence a diffusion process especially its diffusion size, i.e. the number of nodes reached. However, we lack foundational understanding of which kind of node pairs are likely to contribute to an actual information diffusion process, i.e. appear in an information diffusion trajectory. Such understanding is essential to explain and control the prevalence of information spread (e.g. which node pairs should be stimulated to contact at what time in order to maximize the prevalence?). The contact frequency between nodes, as typically used in static networks, is not the only factor that would affect the appearance of a node pairs in an information diffusion trajectory, as we need to consider the time stamps of the contacts as well~\cite{yang2012epidemic, chu2009epidemic, pfitzner2013betweenness,li2017reconstruction}. For instance, the node pairs with a lot of contacts that only happened before the information starts to diffuse are of no importance for the diffusion process.

In this paper, we address the question of which kind of node pairs are likely to contribute to the diffusion of information, considering the $SI$ diffusion process as a start. Specifically, we explore how the probability that a node pair appears in a diffusion trajectory is related to local temporal connection features of the two nodes.
First, we propose the construction of an \emph{information diffusion backbone} $G_{B}(\beta)$ for a $SI$ spreading process with an infection probability $\beta$ on a given temporal network. The construction is based on a large number of information diffusion trajectories. The resultant backbone is a weighted network where the weight of each node pair indicates how likely the node pair contributes to a diffusion process that starts from an arbitrary node. We consider a large number of empirical temporal networks. For each network, we construct diffusion backbones for diverse infection probabilities and study the relationship between these backbones.  We find that backbone topology varies from $G_{B}(\beta=0)$ (which equals the integrated weighted network) when the spreading probability $\beta$ is small to $G_{B}(\beta=1)$ when the infection probability is large. The difference between the two extreme  backbones $G_{B}(\beta=0)$ and $G_{B}(\beta=1)$, suggests the extent to which the backbones with diverse infection rates may vary. The computational complexity of $G_{B}(\beta=0)$ is high. Hence, we investigate further which local connection feature of a node pair may predict the links and the links with a high weight in the backbone $G_{B}(\beta=1 )$. One of the features that we proposed incorporates the time stamps when contacts occur between a node pair. It outperforms other classic features of a node pair derived from the integrated network, which points out the importance of temporal information in determining the role of a node pair in a diffusion process.

The paper is organized as follows. After introducing how to represent a temporal network in Section 2, we explain in Section \ref{Sec:Diffusion Backbone Definition} the process of constructing the information diffusion backbone for a $SI$ diffusion process on a temporal network. We consider a set of empirical temporal networks, which are described in Section \ref{Sec:Empirical Networks}. In Section 5, we present our comparative analysis of the constructed backbones for different infection probabilities and for different networks. In Section 6, we evaluate which local connection features of a node pair, including the measures we proposed, could well predict whether the node pair will be connected in the backbone $G_{B}(\beta=1)$ and with a high weight or not. A discussion in Section 7 concludes the paper.

\section{Representation of a temporal network}
A temporal network can be measured by observing the contacts between each node pair at each time step within a given time window $[0, T]$ and represented as $\mathcal{G}=(\mathcal{N}, \mathcal{L})$. Here, $\mathcal{N}$ is the node set, with the size $N=|\mathcal{N}|$ representing the number of nodes in the network, and $\mathcal{L}=\{l(j, k, t), t\in[0, T]\}$ is the contact set, where the element $l(j, k, t)$ indicates that the nodes $j$ and $k$ have a contact at time step $t$. A temporal network can also be described by a three-dimensional binary adjacency matrix $\mathcal{A}_{N \times N \times T}$, where the elements $\mathcal{A}(j, k, t)=1$ and $\mathcal{A}(j, k, t)=0$ represent, respectively, that there is a contact or no contact between the nodes $j$ and $k$ at time step $t$.

\begin{figure*}[!ht]
\centering
\includegraphics[width=16cm]{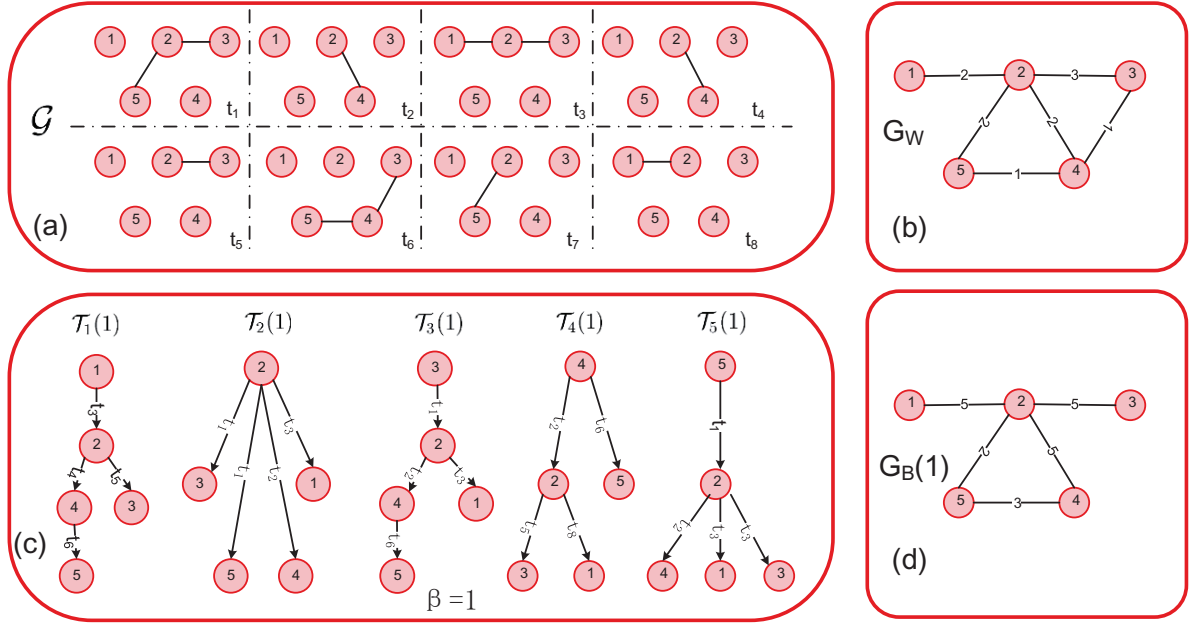}
\caption{\label{Fig:1} (a) A temporal network $\mathcal{G}$ with $N=5$ nodes and $T=8$ time steps.
(b) The integrated weighted network $G_{W}$, in which a link exists between a node pair in $G_{W}$ as long as there is at least one contact between them in $\mathcal{G}$. The weight of a link in $G_{W}$ is the number of contacts between the two nodes in  $\mathcal{G}$.
(c) Diffusion path tree $\mathcal{T}_{i}(\beta)$, where node $i$ is the seed and infection rate is $\beta=1$.
(d) Diffusion backbone $G_{B}(1)$, where the infection probability $\beta=1$ in the $SI$ diffusion process.
The weight on the node pair represents the number of times it appears in all the diffusion path trees.}
\end{figure*}

An integrated weighted network $G_{W}=(\mathcal{N}, \mathcal{L}_{W})$ can be derived from a temporal network $\mathcal{G}$ by aggregating the contacts between nodes over the entire observation time window $T$. In other words, two nodes are connected in $G_{W}$ if there is at least one contact between them in $\mathcal{G}$. Each link $l(j, k)$ in $\mathcal{L}_{W}$ is associated with a weight $w_{jk}$ counting the total number of contacts between node $j$ and $k$ in $\mathcal{G}$. The integrated weighted network $G_{W}$ can therefore be described by a weighted adjacency matrix $A_{N \times N}$, with its element

\begin{equation}
\label{Eq:weight matrix}
A(j, k) = \sum_{t=1}^T \mathcal{A}(j, k, t)
\end{equation}
counting the number of contacts between a node pair.
An example of a temporal network $\mathcal{G}$ and its integrated weighted network $G_{W}$ are given in Figure~\ref{Fig:1}(a) and (b), respectively.

\section{Information Diffusion Backbone}
\label{Sec:Diffusion Backbone Definition}
We propose to characterize how node pairs are involved in diffusion processes by constructing information diffusion backbones. We will construct a backbone for the $SI$ diffusion process with a given infection probability $\beta$ on a temporal network defined above.
We start with the simple case when $\beta=1$. At time step $t=0$, the seed node $i$ is infected and all the other nodes are susceptible. The trajectory of the $SI$ diffusion on $\mathcal{G}$ can be recorded by a \emph{diffusion path tree} $\mathcal{T}_{i}(\beta)$. The diffusion path tree $\mathcal{T}_{i}(\beta)$ records the union of contacts, via which information diffuses. We define the diffusion backbone $G_{B}(\beta)=(\mathcal{N}, \mathcal{L}_{B}(\beta))$ as the union of all diffusion path trees, i.e., $\bigcup\limits_{i=1}^{N}$ $\mathcal{T}_{i}(\beta)$, that start at each node as the seed node. The node set of $G_{B}(\beta)$ is $\mathcal{N}$, and nodes are connected in $G_{B}(\beta)$ if they are connected in any diffusion path tree. Each link in $\mathcal{L}_{B}(\beta)$ is associated with a weight $w_{jk}^{B}$, which denotes the number of times node pair $(j, k)$ appears in all diffusion path trees. An example of how to construct the diffusion backbone is given in Figure~\ref{Fig:1}(c) and (d) for $\beta =1 $. The ratio $\frac{w_{jk}^{B}}{N}$ indicates the probability that the node pair $(j, k)$ appears in a diffusion trajectory starting from an arbitrary seed node.

When $0<\beta<1$, the diffusion process is stochastic. In this case, the backbone can be obtained as the average of a number of realizations of the backbones. Per realization, we run the $SI$ process starting from each node serving as the seed for information diffusion, obtain the diffusion path trees and construct one realization of the diffusion backbone. The weight $w_{jk}^{B}$ of a link in $G_{B}(\beta)$ is the average weight of this link over the $h$ realizations. The computational complexity of constructing $G_{B}(\beta)$ is $\mathcal{O}(N^{3}Th)$, where $T$ is the length of the observation time window of the temporal network.

\section{Empirical Networks}
\label{Sec:Empirical Networks}
\subsection{Description and basic features}
For the construction and analysis of diffusion backbones, we consider a large number of temporal networks that capture two types of contacts, i.e., physical and virtual contacts.
We collect the datasets $Reality$ $mining$ ~\cite{konect:2017:mit, konect:eagle06}, $Hypertext$ $2009$ ~\cite{konect:2017:sociopatterns-hypertext, konect:sociopatterns}, $High$ $School$ $2011$~\cite{fournet2014contact}, $High$ $School$ $2012$~\cite{fournet2014contact}, $High$ $School$ $2013$~\cite{mastrandrea2015contact}, $Primary$ $School$ ~\cite{stehle2011high}, $Workplace$~\cite{genois2015data}, $Haggle$ ~\cite{konect:2017:contact, konect:chaintreau07} and $Infectious$~\cite{isella2011s} that record the face-to-face physical contacts of individuals at MIT, ACM Hypertext 2009 conference, a high school, a primary school, a workplace and the Science Gallery, respectively. We also consider virtual contact datasets recording the mailing and message behavior, including $Manufacturing$ $Email$ ~\cite{konect:2017:radoslaw_email, konect:radoslaw}, $Email$ $Eu$ ~\cite{leskovec2007graph}, $DNC$ $Email$ ~\cite{konect:2017:dnc-temporalGraph} and $Collegemsg$~\cite{panzarasa2009patterns}. The list of the datasets used and their detailed statistics are given in Table~\ref{TB:1}. We consider only the temporal network topologies measured at discrete time steps in these datasets, whereas the during of a time step differ among these datasets. We have removed the time steps without any contact in order to consider the steps that are relevant for information diffusion and to avoid the periods that have no contact due to technical errors in measurements.

\begin{table*}[!ht]\footnotesize
\centering
\caption{\label{TB:1} Basic features of the empirical networks. The number of nodes ($N$), the original length of the observation time window ($T$ in number of steps), the total number of contacts ($|\mathcal{C}|$), the number of links in $G_{W}$ ($|\mathcal{L}_{W}|$) and contact type are shown. }
\newcommand{\minitab}[2][1]{\begin{tabular}{#1}#2\end{tabular}}
\begin{tabular}{ccccccccc}
\hline
\multirow{2}*{Network} & \multirow{2}*{$N$} & \multirow{2}*{$T$} & \multirow{2}*{$|\mathcal{C}|$} & \multirow{2}*{$|\mathcal{L}_{W}|$} &  \multirow{2}*{\minitab[c]{$Contact$ \\ $Type$}}\\\\\hline
Reality Mining (RM)& 96 & 33,452 & 1,086,404 & 2,539  &Physical\\
Hypertext 2009 (HT2009)& 113 & 5,246 & 20,818 & 2,196  &Physical\\
High School 2011 (HS2011)& 126 & 5,609 & 28,561 & 1,710 &Physical\\
High School 2012 (HS2012) & 180 & 11,273 & 45,047 & 2,220 &Physical\\
High School 2013 (HS2013) & 327 & 7,375 & 188,508 & 5,818 &Physical\\
Primary School (PS) & 242 & 3,100 & 125,773 & 8,317 &Physical\\
Workplace (WP)& 92 & 7,104 & 9,827 & 755 &Physical\\
Manufacturing Email (ME)& 167 & 57,791 & 82,876 & 3,250 &Virtual\\
Email Eu (EEU)& 986 & 207,880 & 332,334 & 16,064 &Virtual\\
Haggle & 274 & 15,662 & 28,244 & 2,124  &Physical\\
Infectious & 410 & 1,392 & 17,298 & 2,765 &Physical\\
DNC Email (DNC) & 1866 & 1,8682 & 37,421 & 4,384 &Virtual\\
Collegemsg & 1899 & 5,8911 & 59,835 & 13,838 &Virtual\\

\hline
\end{tabular}
\end{table*}
\subsection{Observation time windows}
We aim to understand which node pair is likely to connected in the backbone, thus contribute to a diffusion process and how such connection in the backbone is related to this node pair's temporal connection features. However, real-world temporal networks are measured for different lengths $T$ of time windows as showing in Table~\ref{TB:1}.  If a diffusion process has a relatively high spreading probability or the temporal network has a relatively long observation time window, almost all the nodes can be reached within a short time. The temporal contacts happened afterwards will not contribute to the diffusion process. Hence, we will select the time windows such that all contacts within each selected time window could possibly contribute, or equivalently, are relevant to a diffusion process. On the other hand, we will consider several time windows for each measured temporal network. This will allow us to understand how the time window of a temporal network may influence the relation between the backbones of different spreading probabilities and relation between a node pair's local connection features and its connection in a backbone.
\begin{figure*}[!ht]
\centering
\includegraphics[width=14cm]{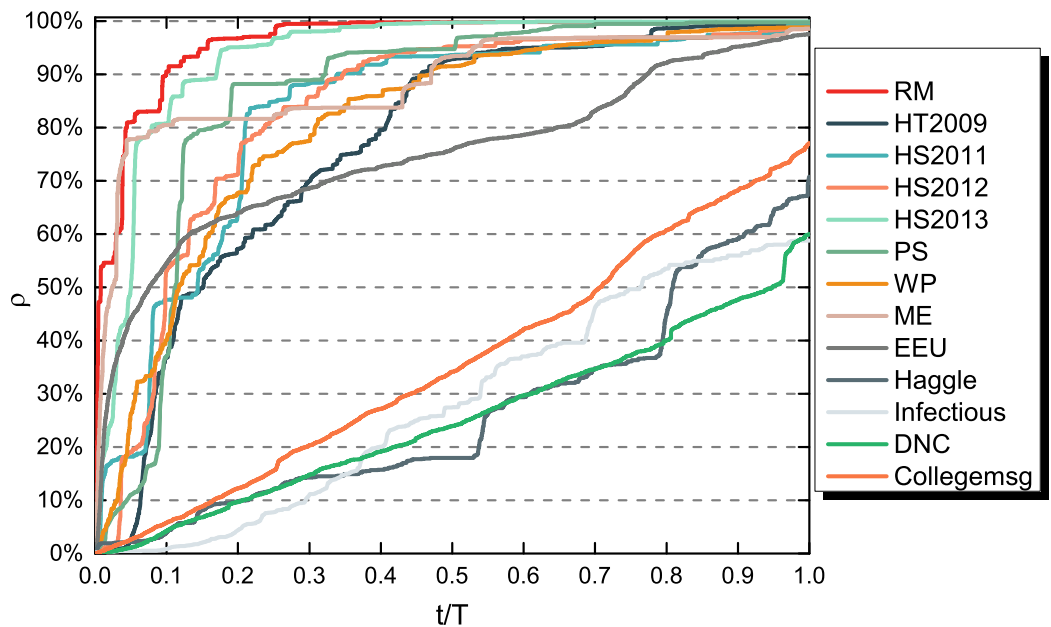}
\caption{\label{Fig:2} Average prevalence $\rho$ of the $SI$ spreading process with $\beta = 1$ on each original empirical temporal network over time. The time steps are normalized by the corresponding observation time window $T$ of each network.}
\end{figure*}
We select the observation time windows for each measured temporal network within its original time window $[0,T]$ as follows. On each measured temporal network with its original observation time window $[0,T]$, we conduct the $SI$ diffusion process with $\beta=1$ by setting each node as the seed of the information diffusion process and plot the average prevalence $\rho$ at each time step, as illustrated in Figure ~\ref{Fig:2}. The time steps are normalized by the original length of observation window $T$.  The average prevalence at the end of the observation $t/T=1$ is recorded as $\rho (t=T)$.  The time to reach the steady state varies significantly across the temporal networks. For networks like $RM$, $HT2009$, the diffusion finishes or stops earlier and contacts happened afterwards are not relevant for the diffusion process. However, the prevalence curves $\rho$ of the last four networks (i.e., $Haggle$, $Infectious$, $DNC$ and $Collegemsg$) increase slowly and continuously over the whole period.

For each real-world temporal network with its original length of observation time window $T$, we consider the following lengths of observation time windows: the time $T_{p\%}$ when the average prevalence reaches $p\%$, where $p \in \{10, 20, \ldots, 90\}$ and $p\%<\rho (t=T)$. For a given measured temporal network $\mathcal{G}=(\mathcal{N}, \mathcal{L})$, we consider maximally $9$ observation time windows. For each length $T_{p\%}$, we construct a sub-temporal network, $\mathcal{G}_{p\%}=(\mathcal{N}, \mathcal{L}_{p\%})$, in which $\mathcal{L}_{p\%}$ include contacts in $\mathcal{L}$ that occur earlier than $T_{p\%}$. The lengths of observation time window $T_{p\%}$ for the empirical networks are shown in Table~\ref{TB:S1} in the \textbf{Supplementary Material}. For a network like $RM$, we can get 9 sub-networks and for network like $Infectious$, we can only obtain 5 sub-networks. In total, 106 sub-networks are obtained. Contacts in all these sub-networks are relevant for SI diffusion processes with any spreading probability $\beta$. Without loss of generality, we will consider all these sub-networks with diverse lengths of observation time windows and temporal network features to study the relationship between diffusion backbones and temporal connection features.

\section{Relationship between Diffusion Backbones}
\label{Sec:Relationship between Diffusion Backbones}

We explore the relationships among the backbones $G_{B}(\beta)$ with different spreading probabilities $\beta \in [0,1]$ on the same temporal network. When the infection probability $\beta \to 0$, the backbone $G_{B}(\beta \to 0)$ approaches the integrated weighted network $G_{W}$. In this case, it takes a long time for the seed node to diffuse the information to another node that it has contacts since the diffusion probability per contact is small. For a temporal network with a finite observation window, the diffusion path tree $\mathcal{T}_{i}(\beta\to 0)$ rooted at the seed node is a star, where the probability that the seed node is connected with another node is proportional to the number of contacts between them. Hence, $G_{B}(\beta=0)\triangleq G_{B}(\beta \to 0)=G_{W}$ except that the weight of each node pair in the two networks are scaled. When the infection probability $\beta$ is small, node pairs with more contacts are more likely to appear in the backbone. The backbone $G_{B}(\beta)$ varies from $G_{B}(0)=G_{W}$ when $\beta\to 0$ to $G_{B}(1)$ when $\beta=1$.
\subsection{Overlap in Links between Backbones}

We investigate first how different these backbones with different spreading probabilities $\beta \in [0,1]$ are and whether $G_{B}(\beta)$ with a small and large $\beta$ can be well approximated by $G_{W}$ and $G_{B}(1)$ respectively.

The similarity between two backbones or two weighted networks in general can be measured by their overlap in links or node pairs with a high weight. For each backbone $G_{B}(\beta)$, links in $\mathcal{L}_{B}(\beta)$ are ordered according to their weights in the backbone in a descending order . Thus the links in the relatively top positions are more likely to be used in the diffusion process. The number of links $|\mathcal{L}_{B}(\beta)|$ in the backbone $G_{B}(\beta)$ decreases as the spreading probability $\beta$ increases, partially reflected in Figure~\ref{Fig:3add} (a) where the number of links in $G_{B}(0)$ and $G_{B}(1)$ are compared. For any backbone with  $\beta \in [0,1]$, we consider the top $|\mathcal{L}_{B}(1)|$ links from $\mathcal{L}_{B}(\beta)$, which are denoted as $\mathcal{L}_{B}^{\ast}(\beta)$. The similarity or overlap between two backbones like $G_{B}(\beta)$ and $G_{B}(\beta=0)$ can be measured by the overlap between $\mathcal{L}_{B}^{\ast}(\beta)$ and $\mathcal{L}_{B}^{\ast}(0)$, defined as
\begin{center}
\label{Eq:overlap}
\begin{equation}
r(\beta,0)=r(\mathcal{L}_{B}^{\ast}(\beta), \mathcal{L}_{B}^{\ast}(0)) = \frac{|\mathcal{L}_{B}^{\ast}(\beta) \cap \mathcal{L}_{B}^{\ast}(0)|}{|\mathcal{L}_{B}(1)|},
\end{equation}
\end{center}
For each temporal network, we construct each backbone $G_{B}(\beta)$, where $\beta = 0.25, 0.5, 0.75, 1$, as the average of 100 iterations of the $SI$ spreading processes starting from each node as the seed,
based on the method illustrated in Section~\ref{Sec:Diffusion Backbone Definition} (The validation that 100 iterations are enough to get a stable backbone is given in Figure~\ref{Fig:iteration_check} in the \textbf{Supplementary Material}). The backbone $G_{B}(\beta=0)$ equals $G_{W}$. The overlap between backbones for dataset $RM$ are shown in Figure~\ref{Fig:3} as an example. More examples are given in Figure~\ref{Fig:S1} in the \textbf{Supplementary Material}). The overlap $r(\beta, 0)$ tends to decrease with the increase of $\beta$ and $G_{B}(\beta=0)$ well approximates the backbones with a small $\beta$. Similarly, $G_{B}(1)$ well approximates the backbones with a large $\beta$. When the observation time window of a temporal network is small, the backbones with different $\beta$ are relatively similar in topology. In this case, a diffusion path tree tends to have a smaller average depth \footnote{The average depth of a tree is the average number of links in the shortest path from the root to another random node in the tree.} and a node pair with a large number of contacts is likely to appear or connect in the backbone, which explains why $G_{W}$ approximates all the backbones including $G_{B}(1)$. These observations motivate us to explore the two extreme backbones $G_{B}(0)$ and $G_{B}(1)$ regarding to how much they differ from or related to each other.

\begin{figure*}[!ht]
\centering
\includegraphics[width=16cm]{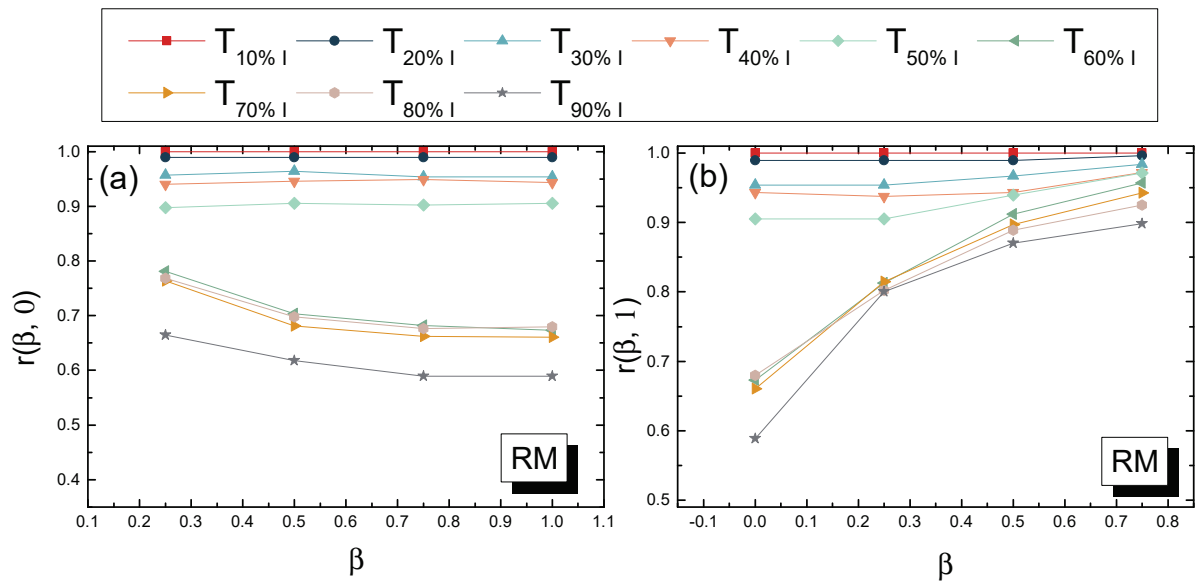}
\caption{\label{Fig:3} (a) Overlap $r(\beta, 0)$ between $G_{B}(\beta)$ and $G_{B}(0)$ as a function of $\beta$ in (sub)networks derived from dataset $RM$; (b) Overlap $r(\beta, 1)$ between $G_{B}(\beta)$ and $G_{B}(1)$ as a function of $\beta$ in (sub)networks derived from dataset $RM$. Diffusion backbones ($0<\beta<1$) are obtained over 100 iterations.}
\end{figure*}

\subsection{Degree of a Node in Different Backbones}
\begin{figure*}[!ht]
\centering
\includegraphics[width=8.5cm]{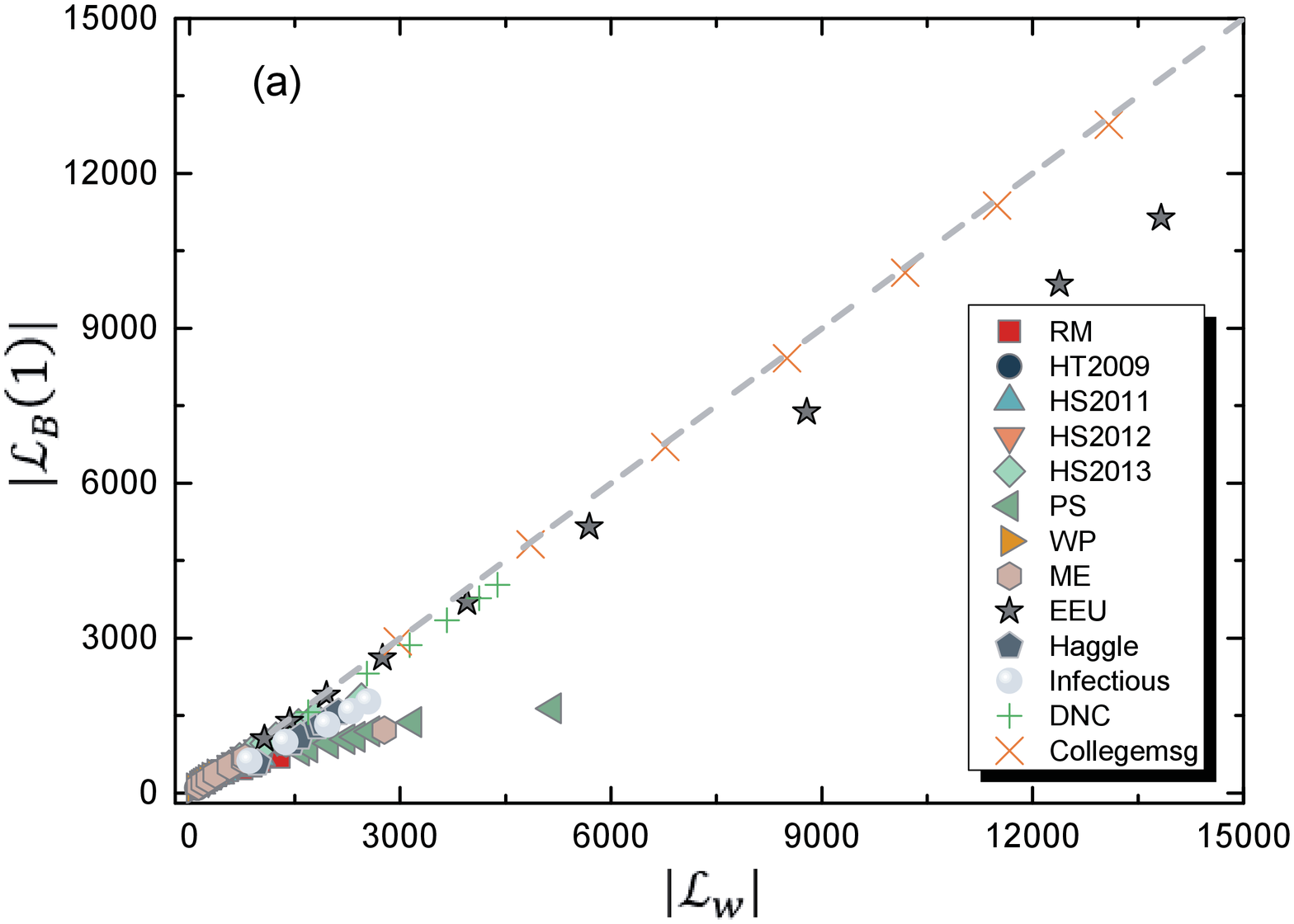}
\includegraphics[width=8.5cm]{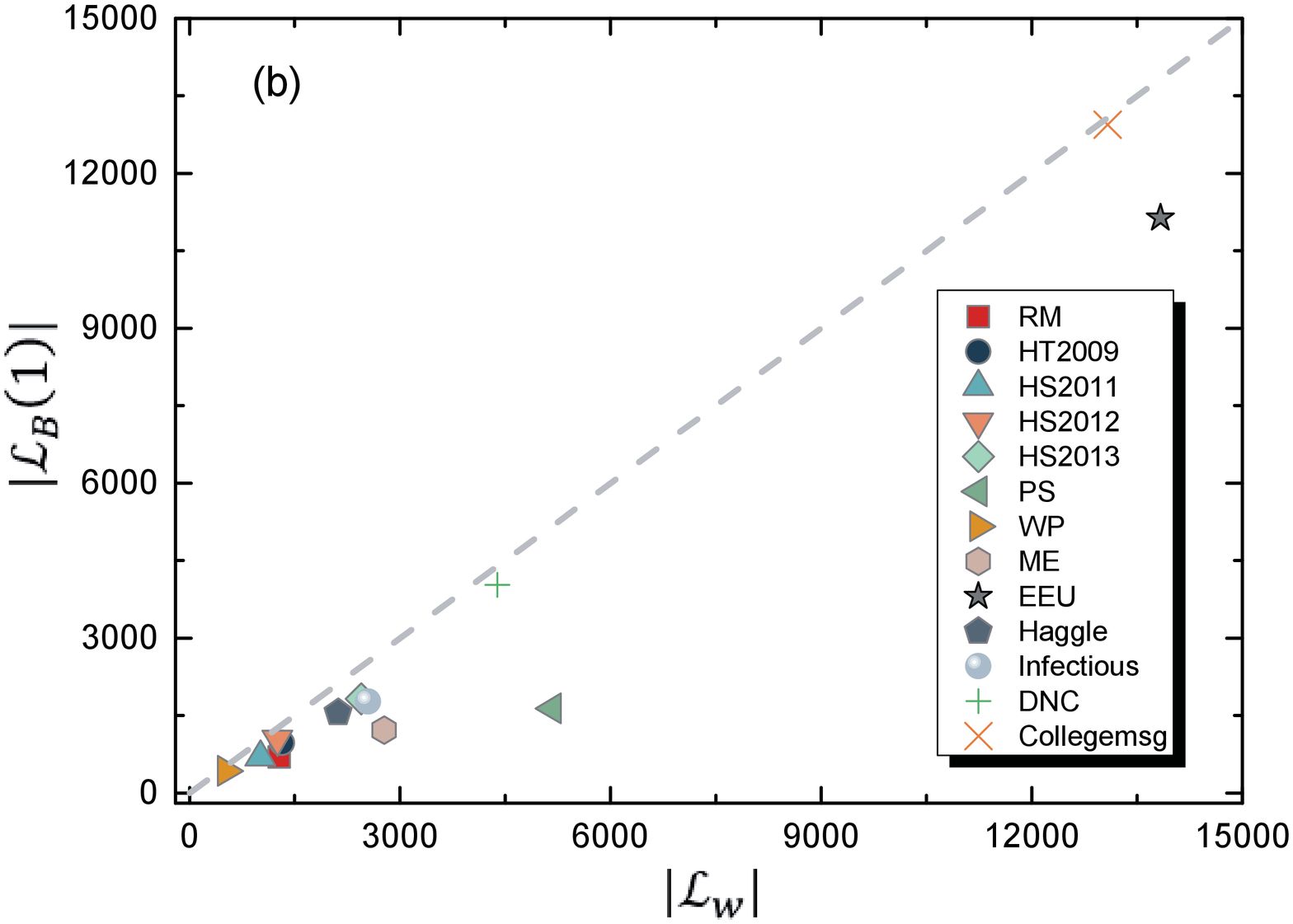}
\caption{\label{Fig:3add} The relationship between the number of links in $G_{W}$ and $G_{B}(1)$ for (a) all the networks with observation windows given in Table~\ref{TB:S1}; (b) the networks with the longest observation windows in each dataset. }
\end{figure*}

From now on, we focus on the two extreme backbones $G_{B}(0)$ = $G_{W}$ and $G_{B}(1)$. A node pair that has contact(s) may not necessarily contribute to a diffusion process. Hence, the degree of a node in $G_{B}(0)$ is larger or equal to its degree in $G_{B}(1)$. The comparison of the number of links in $G_{B}(0)$ and $G_{B}(1)$ in Figure~\ref{Fig:3add} shows that $G_B(1)$ indeed has less links than $G_B(0)$, especially when the observation time window is large. As explained earlier, $G_B(1)$ and $G_B(0)$ are similar to each other in topology when the observation time window is small.

\begin{figure*}[!ht]
\centering
\includegraphics[width=6cm]{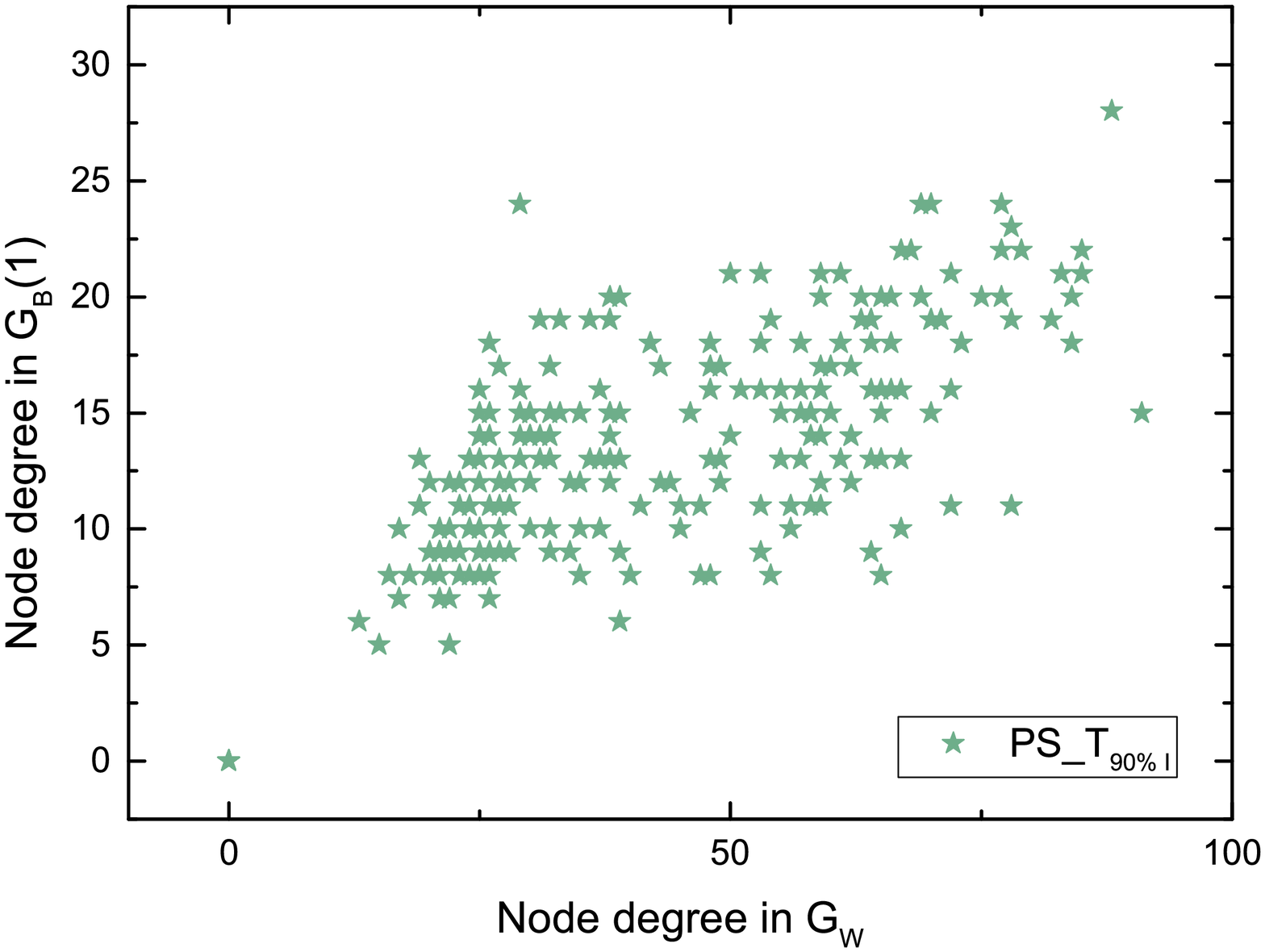}
\includegraphics[width=6cm]{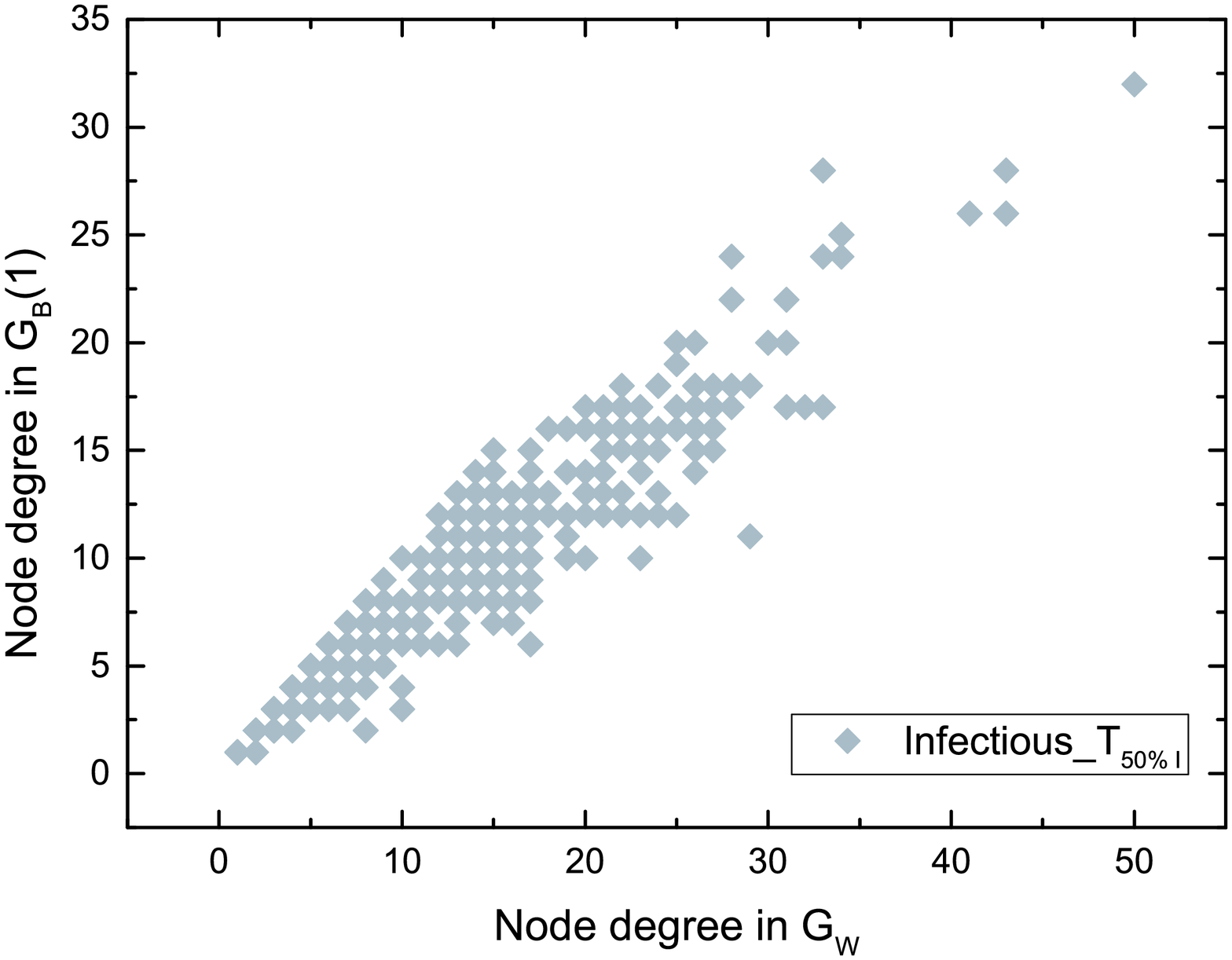}
\caption{\label{Fig:degree correlationRM} Degree correlation between $G_{W}$ and $G_{B}(1)$ for networks $PS$ and $Infectious$ with the longest observation window respectively.  }
\end{figure*}

Furthermore, we explore the degree of a node in $G_{W}=G_{B}(0)$ and $G_{B}(1)$ respectively. Interestingly, a universal finding is that the degree of a node in these two backbones tend to be linearly and positively correlated in all the empirical networks. Table~\ref{TB:degree correlation} in the \textbf{Supplementary Material} provides the Pearson correlation coefficient between the degree of a node in $G_{W}$ and in $G_{B}(1)$ for all the networks, which is above 0.7 for all the networks. Since the topology of $G_{B}(1)$ is a subgraph of $G_{W}$, the degrees of a node in these two networks tend to be linearly correlated if these two networks have a similar number of links. This explains the high degree correlation when the temporal networks have a short observation window. Figure~\ref{Fig:degree correlationRM} shows the scatter plot of the degree of each node in $G_{W}$ and $G_{B}(1)$ respectively for the network with the longest observation window when their backbones $G_{W}$ and $G_{B}(1)$ differ much in the number of links derived from two datasets respectively. The strong degree correlation in all these cases suggests that a node with a high degree in $G_{W}$ tends to have a high degree in $G_{B}(1)$. A node that has contacts with many others tends to be able to propagate the information directly to many others.

\begin{figure*}[!ht]
\centering
\includegraphics[width=16cm]{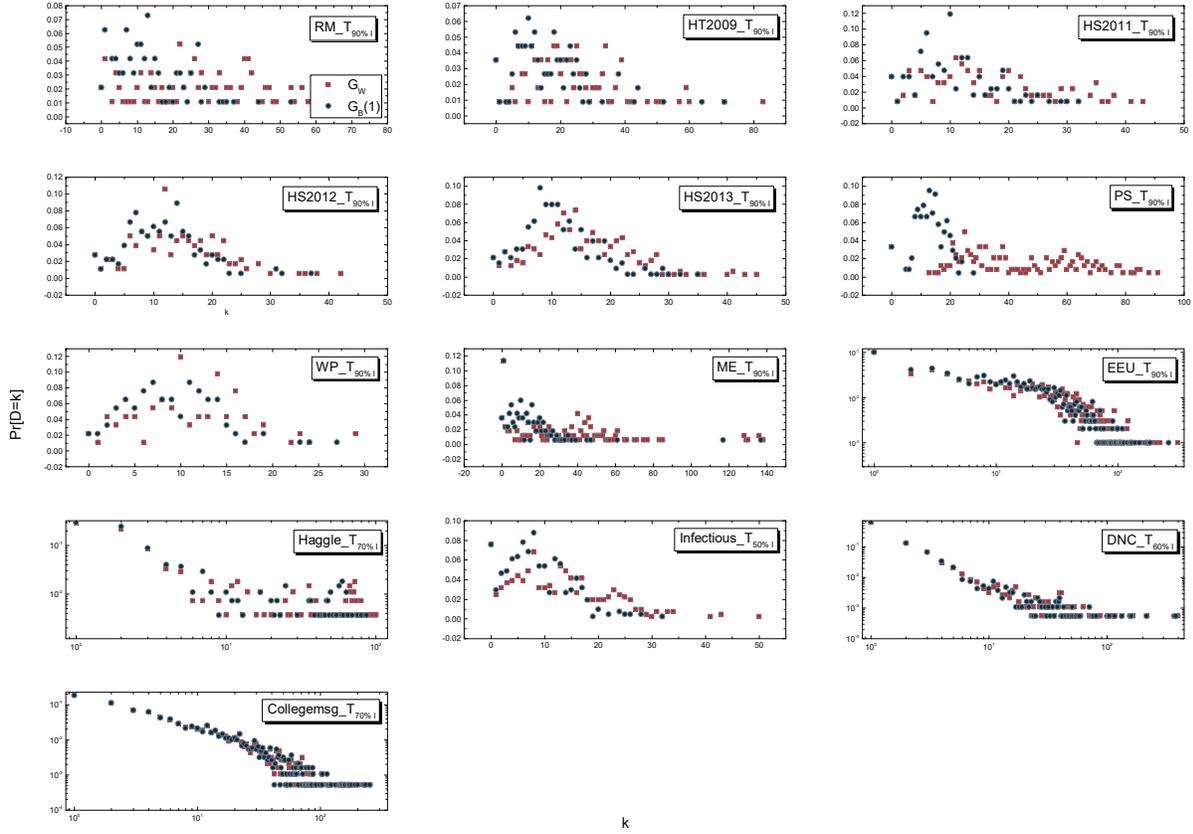}
\caption{\label{Fig:degree distribution} Degree distribution of $G_{W}$ and $G_{B}(1)$ for empirical networks with longest observation window.  }
\end{figure*}

Is this because the degree distribution in $G_{W}$ is highly heterogenous that overrules the temporal orders of the contacts in determining how many other nodes a node is able to reach directly? Figure~\ref{Fig:degree distribution} shows the degree distributions in $G_{W}$ and $G_{B}(1)$ respectively for each temporal network dataset with its longest observation window as given in Table~\ref{TB:S1} when these two backbones differ the most. We find that the degree distributions in these two backbones respectively indeed share a similar shape, which again support the strong linear correlation between the degrees of a node in these two backbones. However, not all networks $G_{W}$ have a power-law degree distribution. The strong degree correlation between $G_{W}$ and $G_{B}(1)$ exists even when $G_{W}$ has a relatively homogeneous degree distribution. This observation motivates us to explore whether a node pair with a high degree product in $G_{W}$ thus also in $G_{B}(1)$ tends to be connected in $G_{B}(1)$ in Section \ref{Sec:Prediction of Diffusion Backbone} .

The degree of a node $j$ in $G_{B}(1)$ tells maximally how many nodes it could propagate the information directly to given that each node is possibly the source of the information, but not necessarily how frequently this node contributes or engages in an information diffusion process when $\beta=1$. The latter is reflected from the node strength of a node in $G_{B}(1)$: $\sum_{k=1}^{N}w_{jk}^{B}(\beta=1)$.

\subsection{Link Weight Variance in Different Backbones}

\begin{figure*}[!ht]
\centering
\includegraphics[width=8.5cm]{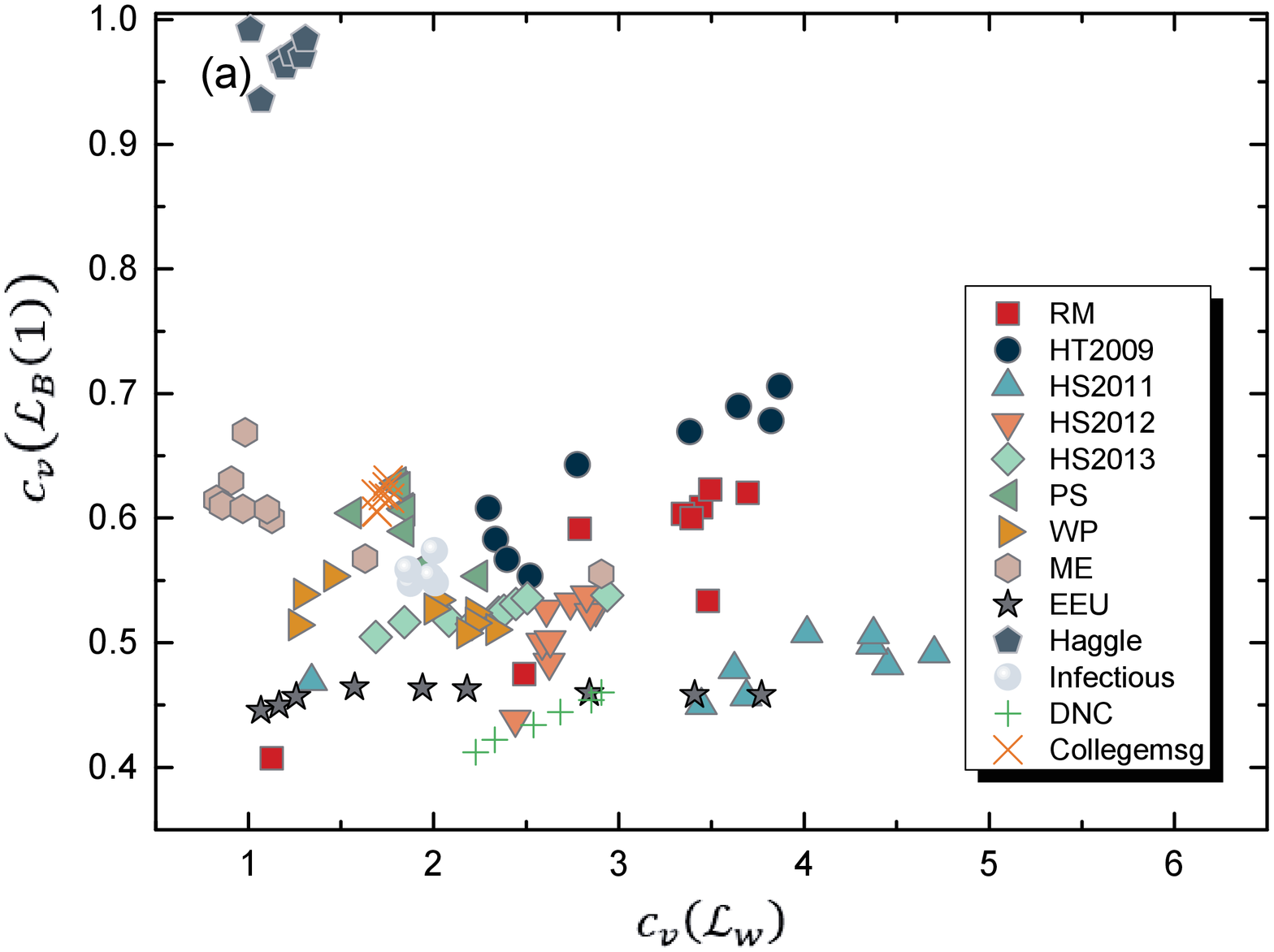}
\includegraphics[width=8.5cm]{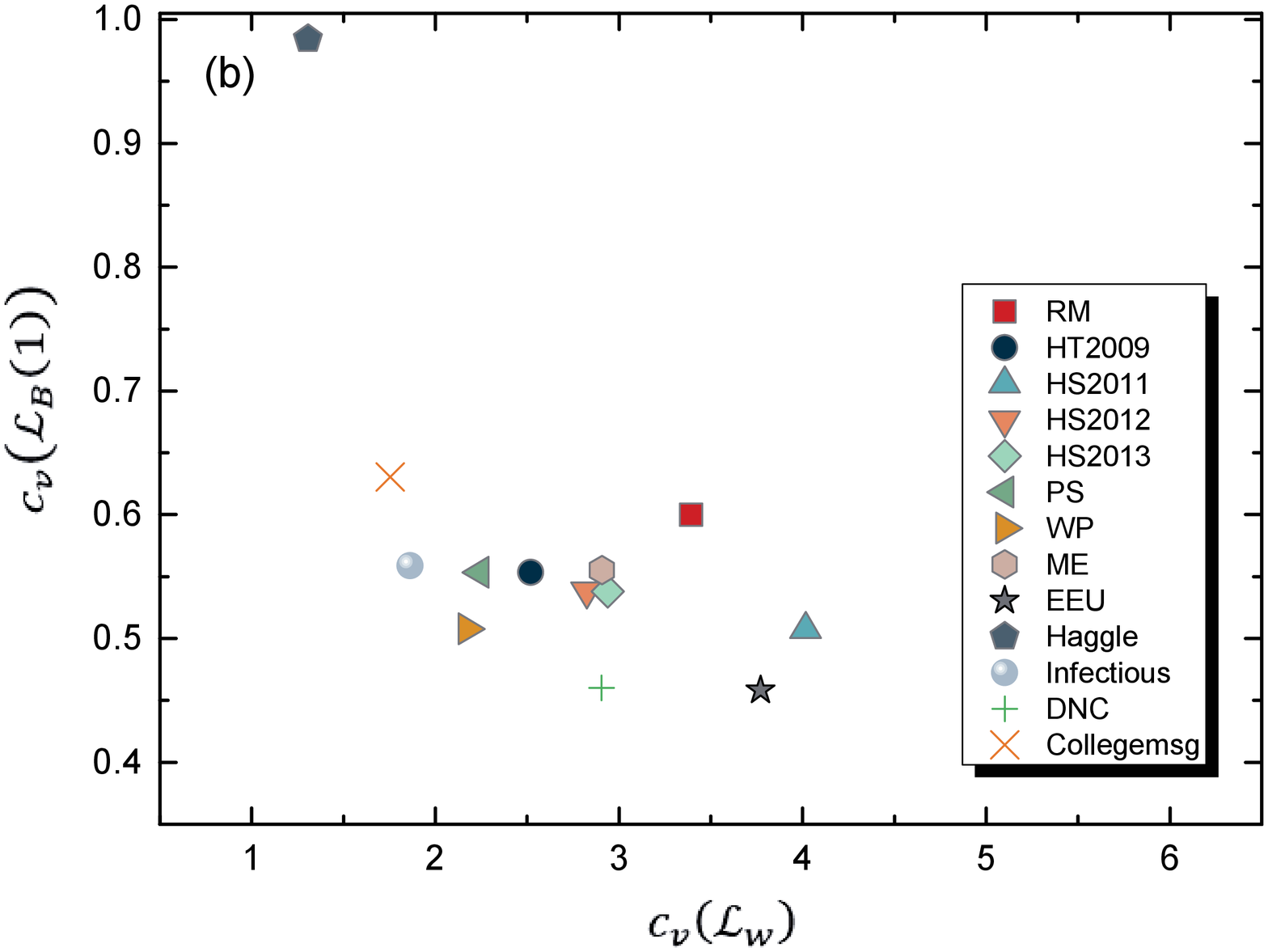}
\caption{\label{Fig:3std}  The relationship between the coefficient of variation $c_{v}$ of the weight distribution in $G_{W}$ and $G_{B}(1)$ for (a) all the networks with observation windows given in Table~\ref{TB:S1}; (b) all the networks with longest observation windows.}
\end{figure*}

The standard deviation of link weights in a backbone indicates how much the links differ in their probability of appearing in a diffusion process. We compare the standard deviation of a link weight normalized by its mean $c_{v}=\frac{\sqrt {\mathrm{Var} [W^{B}]}}{\mathrm{E}[W^{B}]}$ (which is called the coefficient of variation) in $G_B(1)$and $G_B(0)$. Figure~\ref{Fig:3std} shows that the link weights in $G_B(0)$ or equivalently $G_{W}$ is more heterogeneous than that in $G_B(1)$ for almost all the networks we considered. The relatively homogenous link weights in $G_B(1)$ implies that predicting which node pairs tend to have a high weight in $G_B(1)$ can be challenging.

\section{Prediction of the Diffusion Backbone $G_{B}(1)$}
\label{Sec:Prediction of Diffusion Backbone}

In this section, we investigate how to identify the (high weight) links in the backbone $G_{B}(1)$, whose computational complexity is high, based on local and temporal connection features of each node pair. The key objective to understand how local and temporal connection features of a node pair are related to whether the node pair is connected in the $G_{B}(1)$.

We propose to consider systematically a set of local temporal features for node pairs and examine whether node pairs having a higher value of each feature/metric tend to be connected in the backbone $G_{B}(1)$. Some of these features are derived from the integrated network $G_{W}$ whereas the feature $\emph{Time-scaled Weight}$ that we will propose encodes also the time stamps of the contacts between a node pair. These node pair features or metrics include:


$\bullet$ $\emph{Time-scaled Weight}$ of a node pair $(j, k)$ is defined as
\begin{equation}
\label{Eq:TW}
\phi_{jk}(\alpha) = \sum_{m=1}^n(\frac{1}{t_{jk}^{(m)}})^{\alpha}
\end{equation}
where $n$ is the total number of contacts between $j$ and $k$ over the given observation window and $t_{jk}^{(i)}$ is the time stamp when the $i-th$ contact occurs and $\alpha$ is the scaling parameter to control the contribution of temporal information. For the node pairs that have no contact, we assume their temporal weights to be zero.
This metric is motivated by the intuition that when each node is set as the seed of the diffusion process at time $t=0$, the contacts that happen earlier have a higher probability to be used for the actual information diffusion, thus appear in $G_{B}(1)$.
When $\alpha=0$, $\phi_{jk}(0)=w_{jk}^{B}(\beta=0)$ degenerates to the weight of the node pair in $G_{W}$. Larger $\alpha$
implies the node pairs with early contacts have a higher time-scaled weight.

$\bullet$ $\emph{Degree Product}$ of a node pair $(j, k)$ refers to $d_{j}(\beta=0)\cdot d_{k}(\beta=0)$ the product of the degrees of $j$ and $k$ in the integrated network $G_{W}$. If two nodes are not connected in $G_{W}$, their degree product is zero. The motivation for this measure is as follows. Given the degree of each node in $G_{B}(1)$ and if the links are randomly placed, the probability that a node pair $(i,j)$ is connected in $G_{B}(1)$ is proportional to $d_{j}(\beta=1)\cdot d_{k}(\beta=1)$. We have observed in Section~\ref{Sec:Relationship between Diffusion Backbones} that the degree of a node in $G_{W}$ and $G_{B}(1)$ are strongly and positively correlated.  Moreover, only node pairs connected in $G_{W}$ are possible to appear or be connected in $G_{B}(1)$. If the connections in $G_{B}(1)$ are random as in the configuration model ~\cite{newman2001random}, node pairs with a high Degree Product $d_{j}(\beta=0)\cdot d_{k}(\beta=0)$ tend to appear in $G_{B}(1)$.

$\bullet$ $\emph{Strength Product}$ of a node pair $(j, k)$ refers to $s_{j}(\beta=0)\cdot s_{k}(\beta=0)$ the product of the strengths of $j$ and $k$ in the integrated network $G_{W}$, where the strength $s_{j}(\beta=0)= \sum_{i\in\mathcal{N}}A(j, i)$ of a node in $G_{W}$ equals the total weight of all the links incident to this node ~\cite{brainweightedmetrics,grady2012robust}. If two nodes are not connected in $G_{W}$, their Strength Product is zero. This measure is an extension of the Degree Product to weighted networks.

$\bullet$ $\emph{Betweenness}$ of a link in $G_{W}$ counts the number of shortest paths between all node pairs that traverse the link. The distance of each link, based on which the shortest path is computed, is considered to be $\frac{1}{w_{jk}^{B}(\beta=0)}$, inversely proportional to its link weight in  $G_{W}$, since a node pair with more contacts tend to propagate information faster \cite{newman2001scientific, weightedbetween}. Node pairs that are not connected in  $G_{W}$ have a betweenness $0$.

\begin{figure*}[!ht]
\centering
\includegraphics[width=16cm]{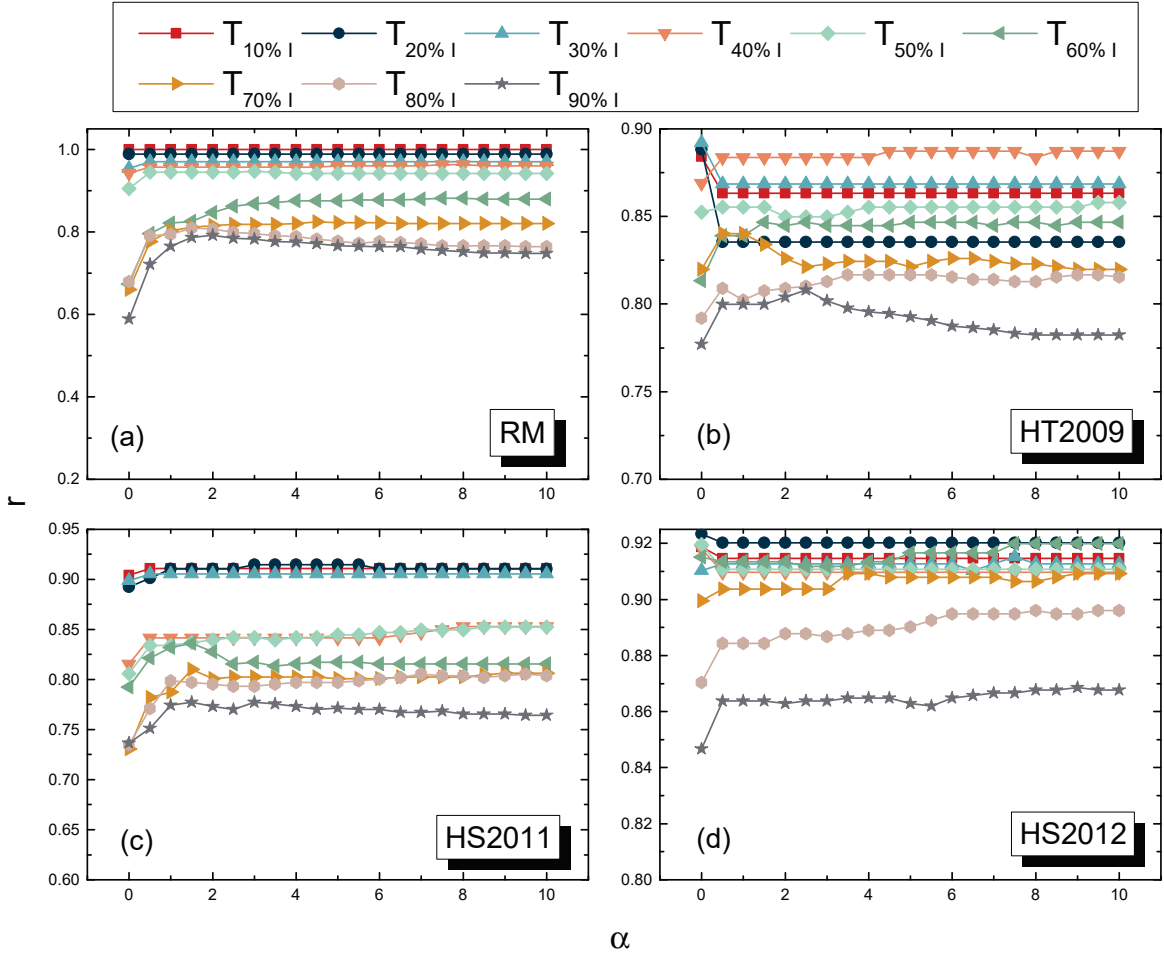}
\caption{\label{Fig:backbone_prediction1} The quality of predicting links in  $G_{B}(1)$ by using the time-scaled weight $\phi_{jk}(\alpha)$ as a function of $\alpha$ in temporal networks derived from datasets (a) $RM$, (b) $HT2009$, (c) $HS2011$ and (d)$HS2012$.  }
\end{figure*}

We explore further whether these node pair features could well predict the connection of node pairs in $G_{B}(1)$. According to the definition of the aforementioned centrality metrics, a higher value of a metric may suggest the connection of the corresponding node pair in $G_{B}(1)$. According to each metric, we rank the node pairs and the $|\mathcal{L}_{B}(1)|$ node pairs with the highest values are predicted as the links in $G_{B}(1)$. The predication quality of a metric, e.g. the time-scaled weight $\phi_{jk}(\alpha)$, is quantified as the overlap $r(\phi_{jk}(\alpha),1)$ between the predicted link set and the link set $\mathcal{L}_{B}(1)$ in $G_{B}(1)$, as defined by Eq. (\ref{Eq:overlap}).

Before we compare all the metrics in their predication powers, we examine first how the scaling parameter $\alpha$ in the time-scaled weight $\phi_{jk}(\alpha)$ influences its predication. Figure~\ref{Fig:backbone_prediction1} and Figure~\ref{Fig:S3} in the \textbf{Supplementary Material} shows that the prediction quality differs mostly when $0\leq\alpha \leq 2$ and remains relatively stable when  $\alpha \geq 2$ in all the temporal networks. Hence, we will confine ourselves to the range $0\leq\alpha \leq 2$.

\begin{figure*}[!ht]
\centering
\includegraphics[width=8.5cm]{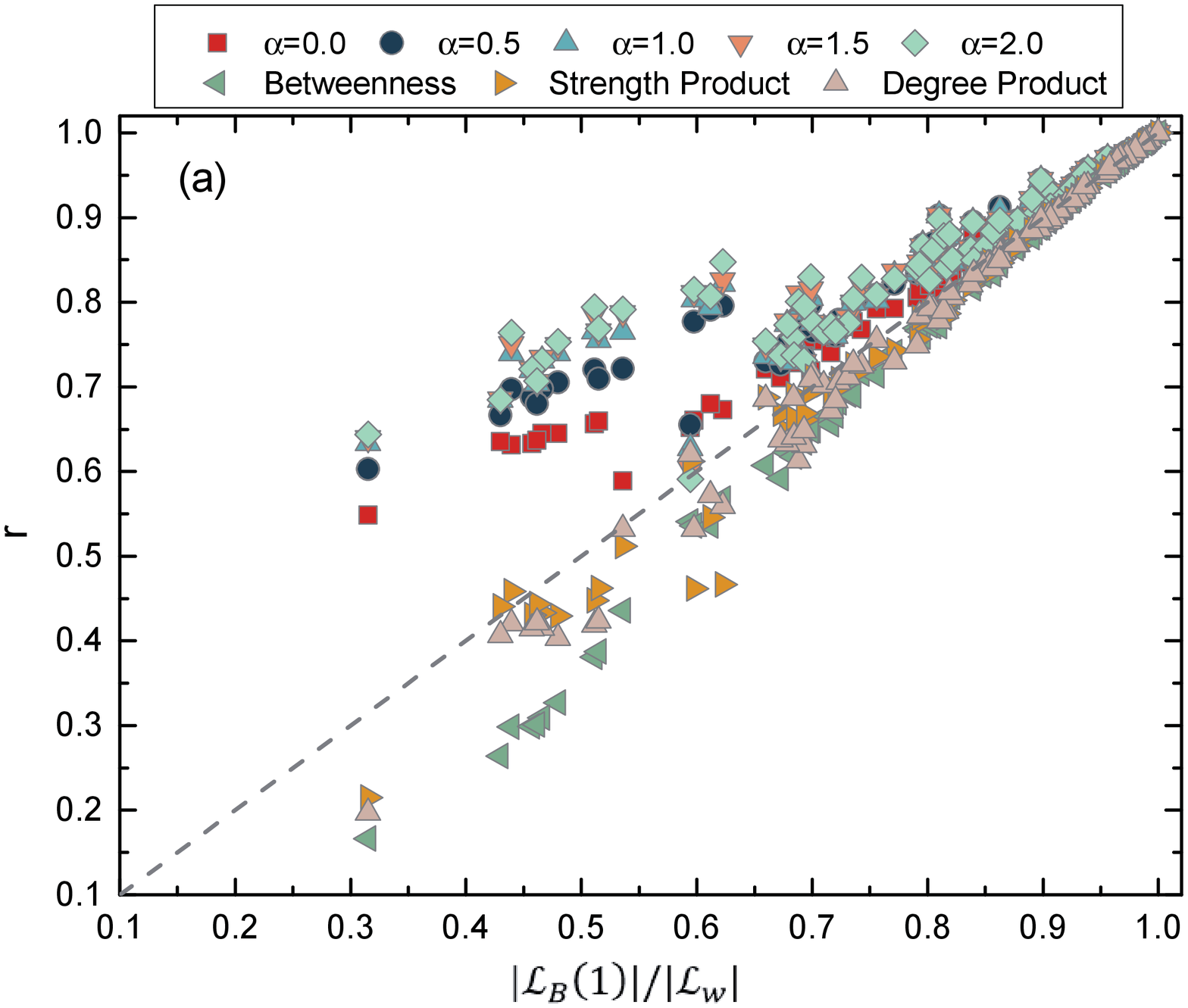}
\includegraphics[width=8.5cm]{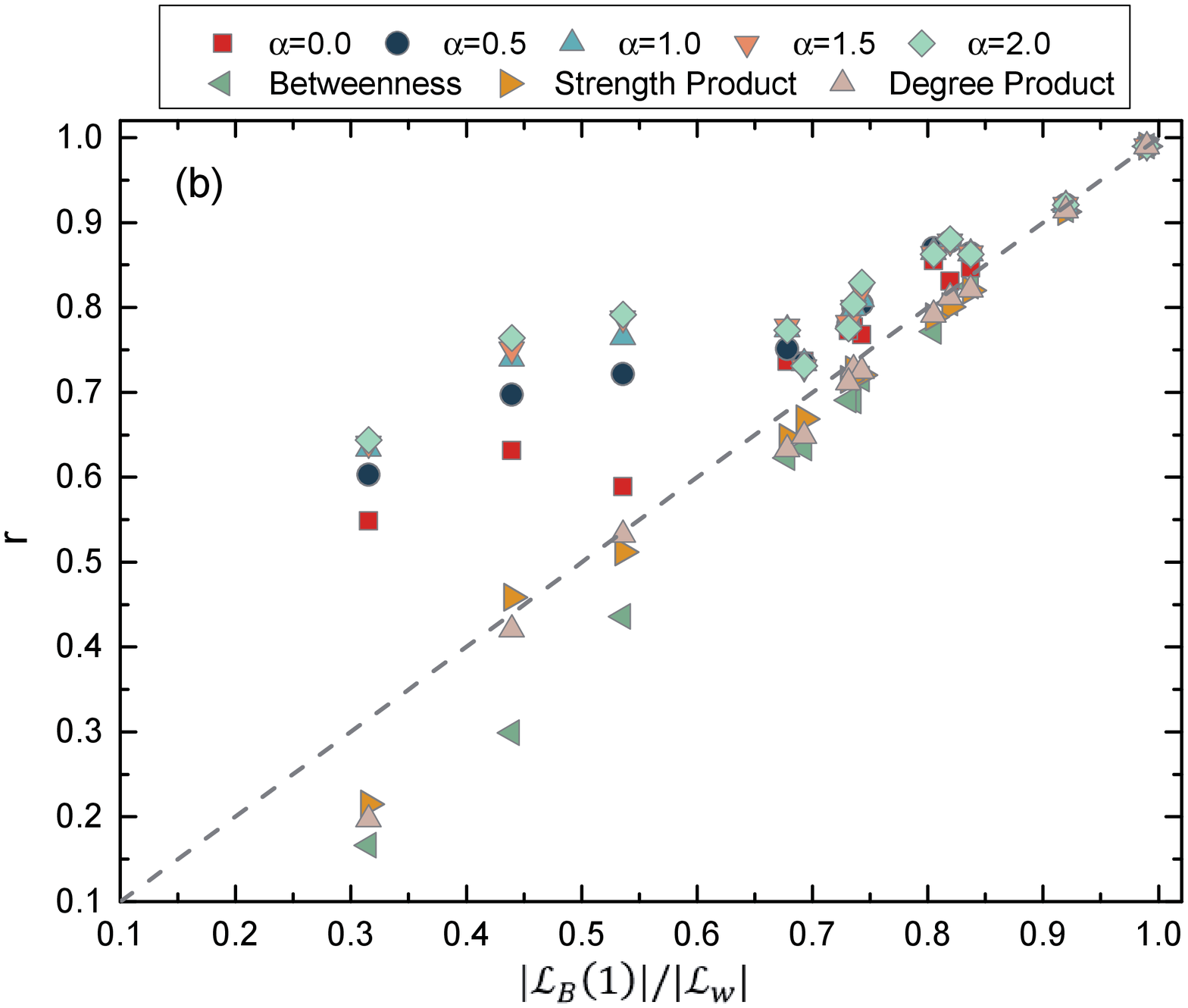}
\caption{\label{Fig:backbone_prediction2} The quality of predicting links in $G_{B}(1)$ by using each metric for (a) all the networks with observation windows given in Table~\ref{TB:S1}; (b) all the networks with longest observation windows. The time-scaled weight with different $\alpha$ values are considered.}
\end{figure*}

The prediction quality $r$ by using each metric versus the ratio $\frac{|\mathcal{L}_{B}(1)|}{|\mathcal{L}_{W}|}|$ of the number of links in $G_{B}(1)$ to that in $G_{W}$ are plotted in Figure~\ref{Fig:backbone_prediction2} for all the empirical temporal networks, with different lengths of the observation time windows. The diagonal curve $r=\frac{|\mathcal{L}_{B}(1)}{|\mathcal{L}_{W}}|$ corresponds to the quality of the random prediction, where $|\mathcal{L}_{B}(1)|$ links are randomly selected from the links in $G_{W}$ as the prediction for the links in $G_{B}(1)$. Degree product, strength product and betweenness perform, in general, worse than or similarly to the random prediction. Even if the connections in $G_{B}(1)$ were random given the degree of each node in $G_{B}(1)$, the  quality $r$ of predicting links in $G_{B}(1)$ by using the degree product is close that of the random prediction, if the distribution of the degree product is relatively homogeneous or if the $\frac{|\mathcal{L}_{B}(1)|}{|\mathcal{L}_{W}|}|$ is large. The degree distribution in $G_{B}(1)$ is indeed relatively homogeneous and $\frac{|\mathcal{L}_{B}(1)|}{|\mathcal{L}_{W}|}|$ is large in most empirical networks. This explains why the degree product performs similarly to the random predication.

The link weight in $G_{W}$, equivalently, $\phi_{jk}(\alpha=0)$, outperforms the random prediction, whereas the time-scaled weight $\phi_{jk}(\alpha)$ with a larger $\alpha$ performs better. Node pairs with many contacts that occur early in time tend to contribute to the actual information propagation, i.e. be connected in $G_{B}(1)$. This observation suggests that the temporal information is essential in determining the role of nodes in a spreading process.

\begin{figure*}[!ht]
\centering
\includegraphics[width=8.5cm]{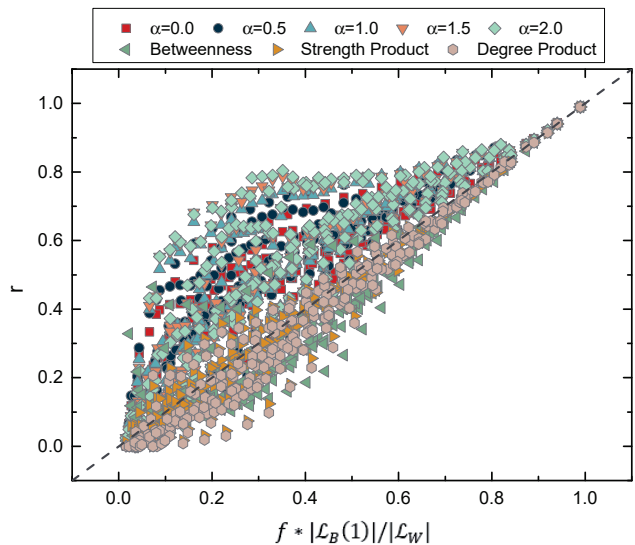}
\caption{\label{Fig:predict_top_links_in_GB1_large_window} The quality $r$ of predicting top weight links in $G_{B}(1)$ by using each metric for all the networks with longest observation windows in each dataset. The time-scaled weight with different $\alpha$ values are considered.}
\end{figure*}

We investigate also whether these metrics can predict the links with the highest weights in $G_{B}(1)$. The quality $r$, as defined earlier, of predicting the top $f$ fraction of links with the highest weight in $G_{B}(1)$ is plotted in Figure~\ref{Fig:predict_top_links_in_GB1_large_window}. We choose the top $f*|\mathcal{L}_{B}(1)|$ node pairs according to each metric as the prediction of the top $f*|\mathcal{L}_{B}(1)|$ links in $G_{B}(1)$ with the highest weights. We consider the networks with the longest observation window from each dataset. The diagonal curve $r=f*\frac{|\mathcal{L}_{B}(1)}{|\mathcal{L}_{W}}|$ corresponds to the quality of the random prediction. Similar to the prediction of all the links in  $G_{B}(1)$, the time-scaled weight $\phi_{jk}(\alpha)$ with a large $\alpha$ performs the best in predicting high weight links in $G_{B}(1)$, addression again the important role of the temporal information of contacts.

\section{Conclusions \& Discussion}
\label{Sec:Conclusion}

Much effort has been devoted to understand how temporal network features influence the prevalence of a diffusion process. In this work, we addressed the further question: node pairs with what kind of local and temporal connection features tend to appear in a diffusion trajectory or path, thus contribute to the actual information diffusion? We consider the Susceptible-Infected spreading process with an infection probability $\beta$ per contact on a temporal network as the starting point. We illustrate how to construct the information diffusion backbone $G_{B}(\beta)$ where the weight of each link tells the probability that a node pair appears in a diffusion process starting from a random node. We unravel how these backbones corresponding to different infection probabilities relate to each other with respect to their topologies (overlap in links), the heterogeneity of the link weight, and the correlation in nodal degree. These relations point out the importance of two extreme backbones: $G_{B}(1)$ and the integrated network $G_{B}(0)=G_{W}$, between which $G_{B}(\beta)$ varies. We find that the temporal node pair feature that we proposed could better predict the links in $G_{B}(1)$ as well as the high weight links than the features derived from the integrated network. This universal finding across all the empirical networks highlights that temporal information are crucial in determining a node pair's role in a diffusion process. A node pair with many early contacts tends to appear in a diffusion process.

This work reminds us the studies a decade ago about the information transportation via the shortest path on a static network. How frequently a link appears in a shortest path thus contributes to the transportation of information is reflected by the weight of the link in the backbone or overlay, the union of shortest paths between all node pairs ~\cite{overlay}. This weight equals the betweenness, which has a high computational complexity, thus motivated the exploration how a node pair's local connection features are related to its betweenness.

The study of information diffusion paths on a temporal network is more complex due to the extra dimension of time. Our finding that early contacts with a quadratic decay in weight over time indicates the appearance of a node pair in a diffusion path, suggests the possibility to predict the appearance of a node pair in a diffusion path in a long period based on its early contacts within a short period, an interesting follow-up question. This work opens new challenging questions like which nodes tend to be reached early and more likely by the information, how such heterogenous features at node or link level are related to local temporal connection features, beyond different spreading models that can be further considered.

\section{Acknowledgments}

This work has been partially supported by the China Scholarship Council (CSC).

\section*{References}

\bibliographystyle{iopart-num}
\bibliography{Bibliography}

\clearpage
\begin{center}
\begin{flushleft}
\vspace{10em}
\begin{center}
\Large{\bf{Supplementary Material for\\}}
\vspace{1em}
\large{\bf{Information diffusion backbones in temporal network}}
\end{center}
\vspace{1em}


\end{flushleft}
\end{center}

\renewcommand
\thefigure{S\arabic{figure}}
\renewcommand
\thetable{S\arabic{table}}
\renewcommand
\theequation{S\arabic{equation}}
\setcounter{figure}{0}
\setcounter{table}{0}
\setcounter{equation}{0}
\renewcommand\refname{Supplementary References}
\renewcommand\figurename{Supplementary Figure}

{\noindent\large{\bf{S1. Data Description}}}


\begin{table*}[!ht]\small
\centering
\caption{\label{TB:S1} The lengths of the observation time window that we choose based on the average prevalence $\rho$ when $\beta = 1$. For instance, $T_{90\%}$ represents the time when the prevalence reaches $\rho = 90\%$. }
\newcommand{\minitab}[2][1]{\begin{tabular}{#1}#2\end{tabular}}
\begin{tabular}{cccccccccc}
\hline
$Network$ & $T_{90\%}$ &$T_{80\%}$ &$T_{70\%}$ &$T_{60\%}$ &$T_{50\%}$ &$T_{40\%}$ &$T_{30\%}$ &$T_{20\%}$ &$T_{10\%}$\\\hline
RM& 3325 & 1482 & 1278 & 987  & 257 &133 &111 &34 &5\\
HT2009& 2394 & 2131 & 1575 & 1154 & 790 & 568 &439 & 377 &332\\
HS2011& 1903 & 1177 & 1152 & 1001 &805 &447 &425 &396 &47\\
HS2012 & 3915 & 2680 & 1907 & 1481 &1109 &1043 &925 &675 &403\\
HS2013 & 1253 & 583 & 406 & 395 &369 &236 &195 &113 &50\\
PS & 997 &510 & 378 & 359 &347 &323 &287 &276 &136\\
WP & 3328 & 2186 & 1538 & 1133 &832 &708 &400 &320 &218\\
ME& 27189 &5096 &1885 &1735 &1387 &731 &461 &285 &168\\
EEU& 160710 &134342 &67883 &27531 &15792 &8100 &4047 &2348 &1490\\
Haggle & / & / &15640 &14229 &12668 &12440 &9523 &8416 &3293\\
Infectious & / & / & / & / & 1062 &955 &751 &553 & 410\\
DNC & / & / & / &18680 &17712 &14918 &11420 &7817 &3860\\
Collegemsg & / & / & 54493 & 46419 & 41663 & 33889 & 26018 & 17367 & 9747\\

\hline
\end{tabular}
\end{table*}

\clearpage
{\noindent\large{\bf{S2. Number of iterations to compute the backbone}}}\\

\begin{figure*}[!ht]
\centering
\includegraphics[width=16cm]{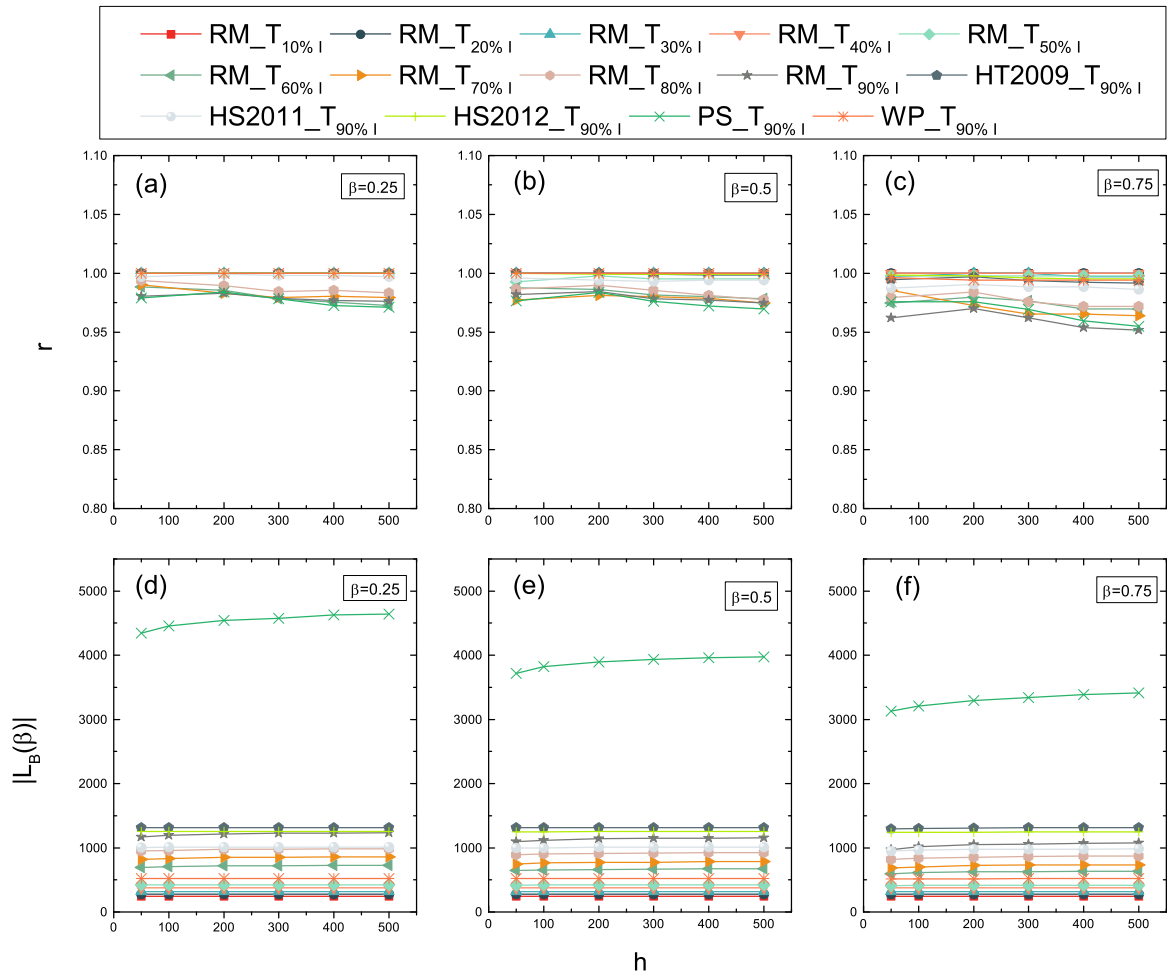}
\caption{\label{Fig:iteration_check} (a-c) Overlap $r$ between backbone obtained from 100 iterations with the backbones obtained from $h=50, 200, 300, 400, 500$ iterations on different temporal networks. (d-f) The number of links in the backbones as a function of the number of iterations.
 }
\end{figure*}

We explore whether 100 iterations is sufficient to get a representative backbone when $0<\beta<1$. Given the temporal network and $\beta$, we first construct the diffusion backbones by choosing the number of iterations as $50, 100, 200, 300, 400, 500$, and then we compute the overlap $r$ between the backbone obtained as the average of 100 iterations with the backbones obtained as the average of $50, 200, 300, 400, 500$ iterations, respectively. The overlap $r$ is defined the same as Eq.~\ref{Eq:overlap}. As the complexity of computing backbones is high, we consider a large number networks but not all. Figure~\ref{Fig:iteration_check} shows the number of links remains relatively unchanged when the number of iterations equals or is above 100. The overlap $r$ is in general high, above $0.95$. These observations support that we could obtain a relatively representative backbone as the average of 100 realizations of the backbone constructions.

\clearpage
{\noindent\large{\bf{S3. Relationship between backbones}}}

\begin{figure*}[!ht]
\centering
\includegraphics[width=13cm]{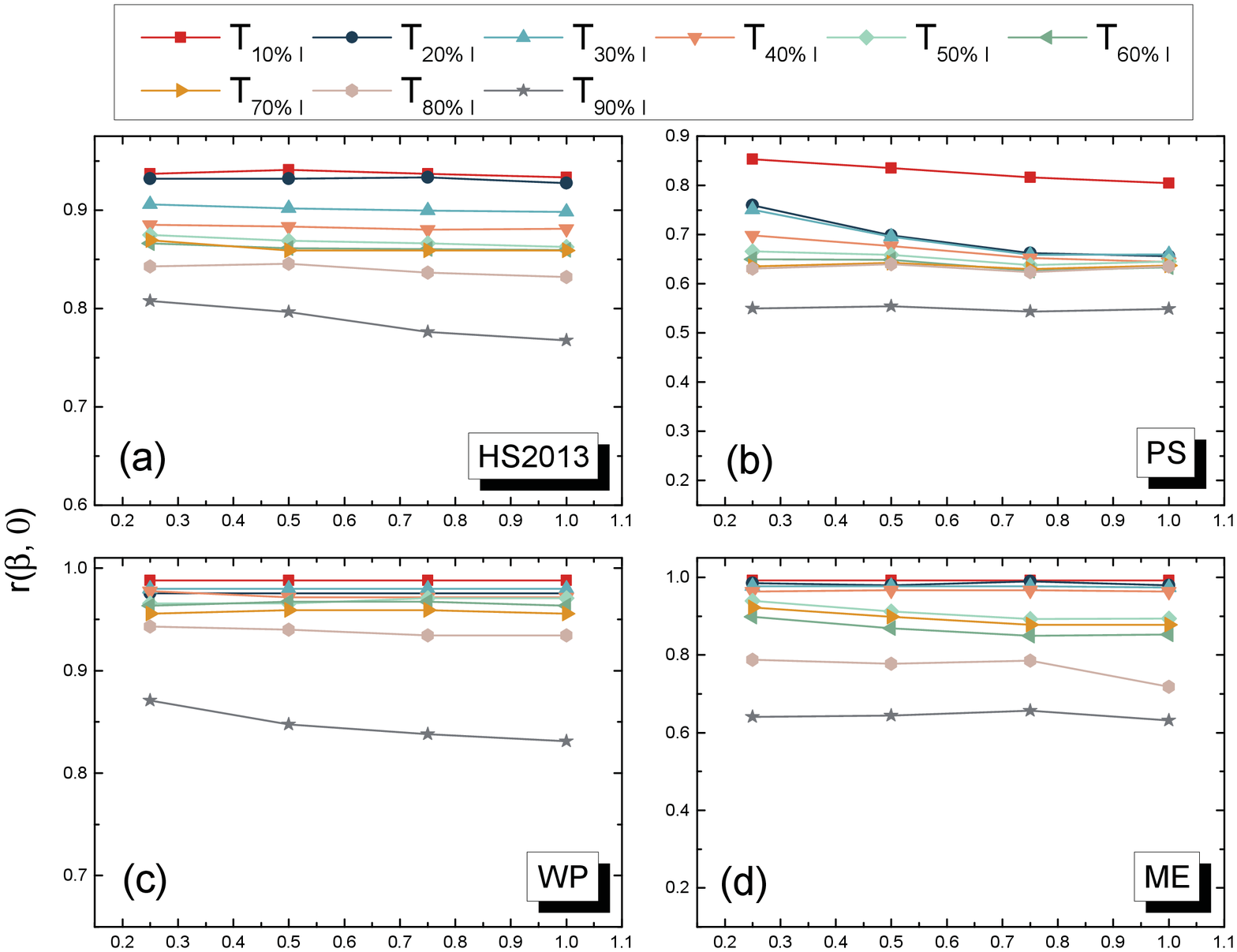}
\includegraphics[width=13cm]{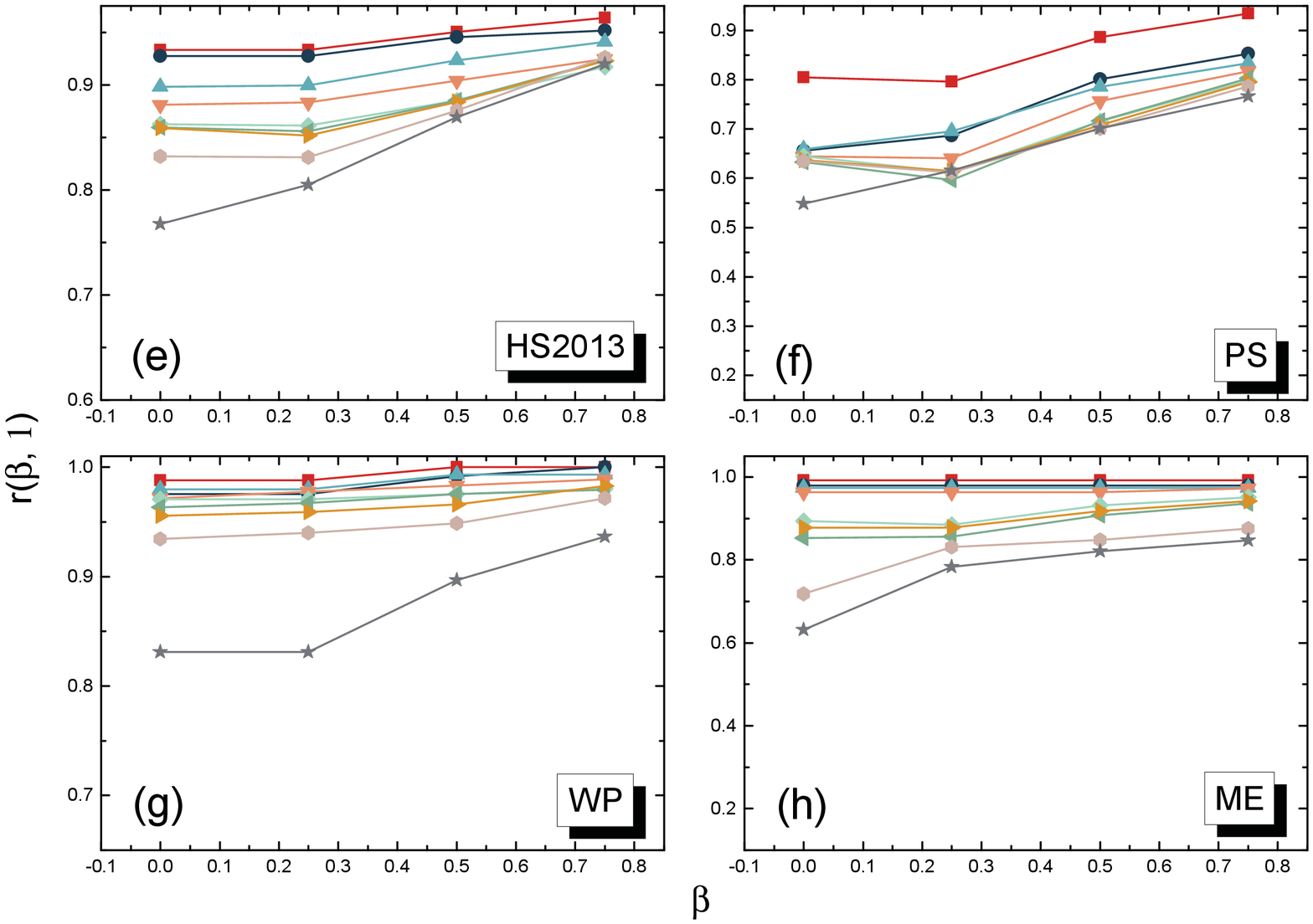}
\caption{\label{Fig:S1} (a-d) Overlap $r(\beta, 0)$ between $G_{B}(\beta)$ and $G_{B}(0)$ as a function of $\beta$ in (sub)networks derived from dataset $HS2013$, $PS$, $WP$ and $ME$; (e-h) Overlap $r(\beta, 1)$ between $G_{B}(\beta)$ and $G_{B}(1)$ as a function of $\beta$ in (sub)networks derived from dataset $HS2013$, $PS$, $WP$ and $ME$. Diffusion backbones ($0<\beta<1$) are obtained from 100 iterations. }
\end{figure*}

\clearpage
{\noindent\large{\bf{S4. Backbone prediction}}}

\begin{figure*}[!ht]
\centering
\includegraphics[width=12cm]{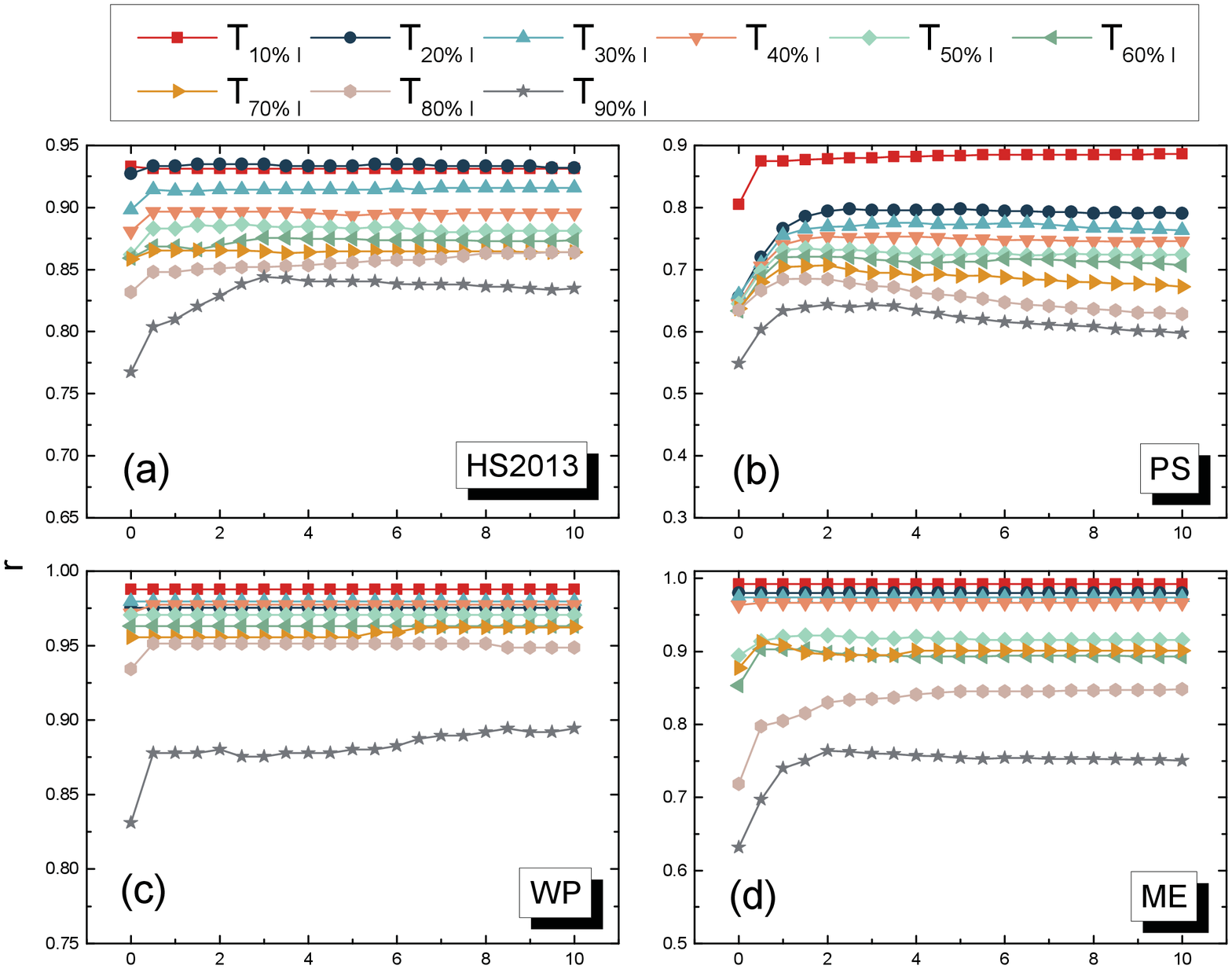}
\includegraphics[width=12cm]{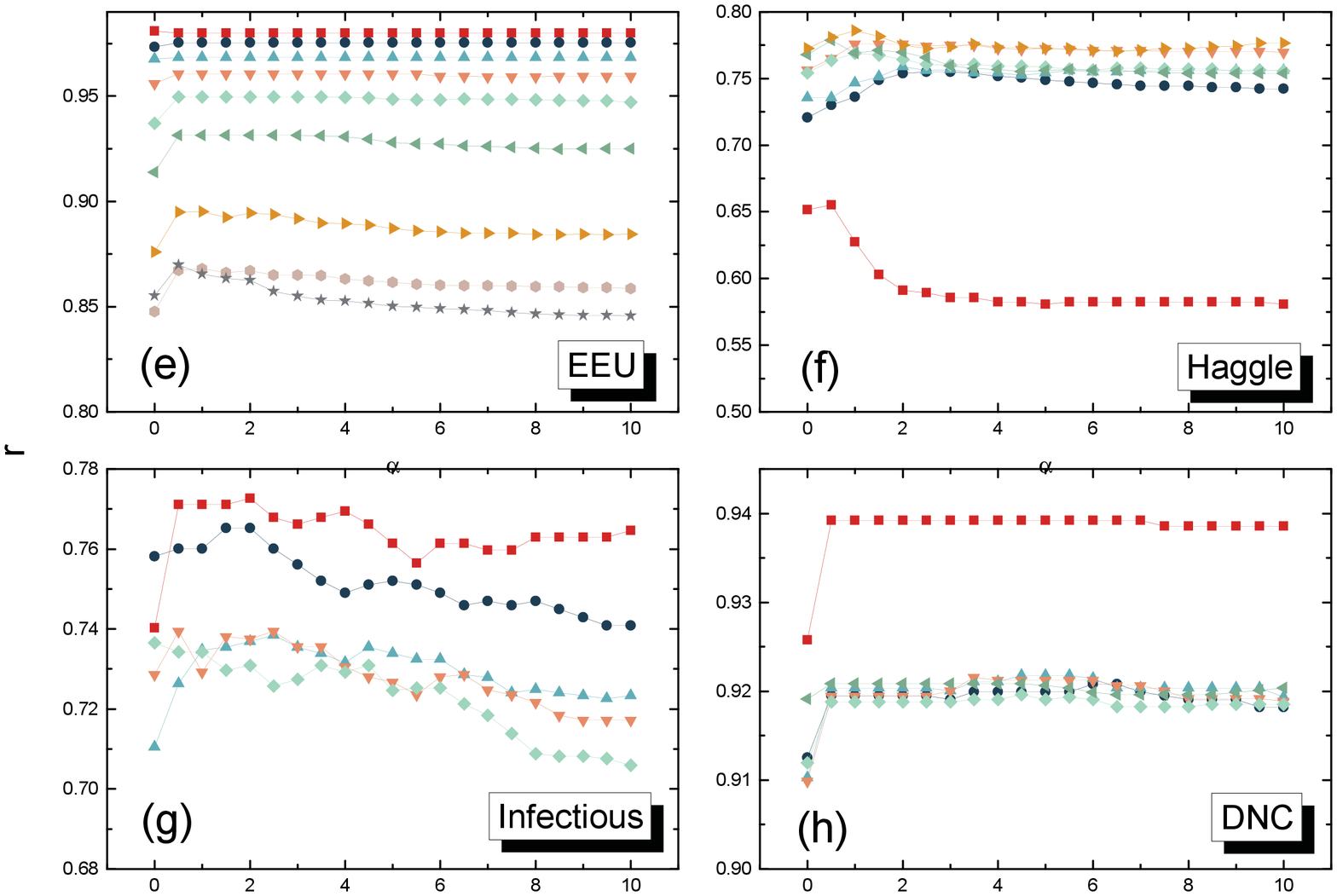}
\includegraphics[width=6cm]{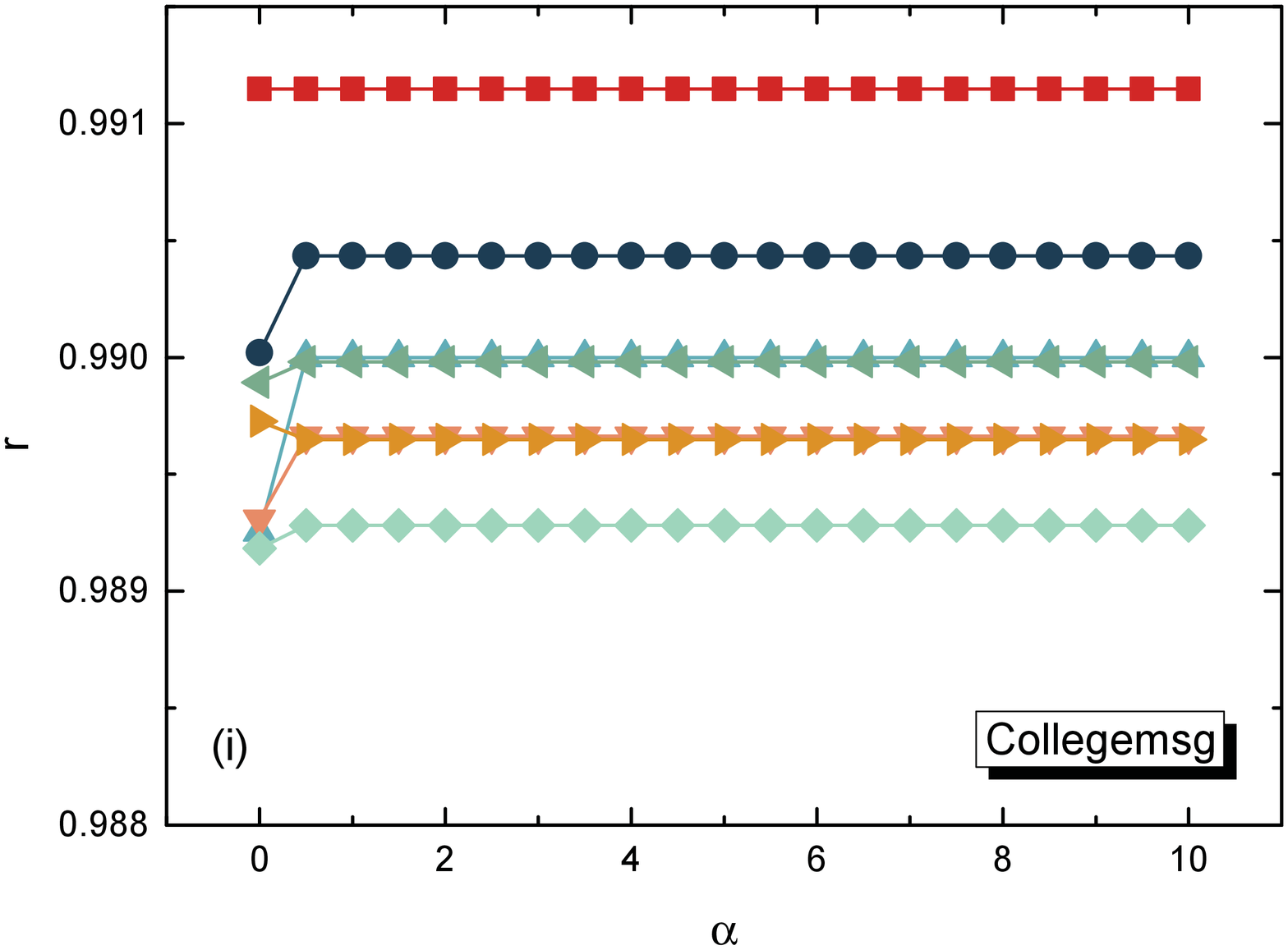}
\caption{\label{Fig:S3} The quality of predicting links in  $G_{B}(1)$ by using the time-scaled weight $\phi_{jk}(\alpha)$ as a function of $\alpha$ in temporal networks derived from datasets (a) $HS2013$; (b) $PS$; (c) $WP$; (d) $ME$; (e)$EEU$; (f) $Haggle$; (g) $Infectious$; (h) $DNC$; (i) $Collegemsg$.}
\end{figure*}

\clearpage
{\noindent\large{\bf{S5. Degree correlation between $G_{W}$ and $G_{B}(1)$}}}\\

\begin{table*}[!ht]\small
\centering
\caption{\label{TB:degree correlation} Pearson correlation coefficient $P(G_{W}, G_{B}(1))$ between node degree in $G_{W}$ and $G_{B}(1)$ in all the networks. }
\newcommand{\minitab}[2][1]{\begin{tabular}{#1}#2\end{tabular}}
\begin{tabular}{cccccccccc}
\hline
$Network$ & $T_{90\%}$ &$T_{80\%}$ &$T_{70\%}$ &$T_{60\%}$ &$T_{50\%}$ &$T_{40\%}$ &$T_{30\%}$ &$T_{20\%}$ &$T_{10\%}$\\\hline
RM&0.8491 &0.8672 &0.8380 &0.8461  &0.9775 &0.9908 &0.9930 &0.9992 &1\\
HT2009&0.9665 &0.9744 &0.9831 &0.9830 &0.9911 &0.9924 &0.9956 &0.9915 &0.9829\\
HS2011&0.9352 &0.9318 &0.9256 &0.9673 &0.9558 &0.9549 &0.9746 &0.9703 &0.9722\\
HS2012 &0.9656 &0.9763 &0.9829 &0.9856 &0.9875 &0.9873 &0.9849 &0.9853 &0.9866\\
HS2013 &0.9368 &0.9634 &0.9717 &0.9739 &0.9747 &0.9791 &0.9784 &0.9865 &0.9857\\
PS &0.7022 &0.7606 &0.7836 &0.7996 &0.8180 &0.8195 &0.8033 &0.7808 &0.9051\\
WP &0.9422 &0.9899 &0.9914 &0.9934 &0.9929 &0.9938 &0.9956 &0.9937 &0.9975\\
ME &0.7624 &0.9442 &0.9892 &0.9265 &0.9877 &0.9967 &0.9977 &0.9981 &0.9990\\
EEU& 0.9913 &0.9909 &0.9920 &0.9936 &0.9936 &0.9965 &0.9968 &0.9965 &0.9967\\
Haggle & / & / &0.9872 &0.9859 &0.9844 &0.9842 &0.9838 &0.9815 &0.9734\\
Infectious & / & / & / & / & 0.9421 &0.9447 &0.9270 &0.9347 & 0.9252\\
DNC & / & / & / &0.9967 &0.9960 &0.9950 &0.9941 &0.9933 &0.9935\\
Collegemsg & / & / & 0.9999 & 0.9999 & 0.9999 & 0.9998 & 0.9998 & 0.9998 & 0.9998\\

\hline
\end{tabular}
\end{table*}

\end{document}